\documentclass[10pt, aip,cha,reprint,onecolumn,nofootinbib]{revtex4-2}
\usepackage{graphicx}
 \graphicspath{{./figures/}}
\usepackage{dcolumn}
\usepackage{amsmath,amsfonts,bm}
\usepackage{physics}
\usepackage{slashed}
\usepackage[format=hang,%
        singlelinecheck=off,%
        justification=raggedright,%
        caption=false]{subfig}
\usepackage[usenames,dvipsnames,svgnames]{xcolor}
\usepackage{colortbl}
\definecolor{Gray}{gray}{0.75}
\usepackage{multirow}
\usepackage{makecell}
\usepackage{hhline}
\usepackage{mathrsfs}
\usepackage[shortlabels,inline]{enumitem}

\let\textcite\relax
\makeatletter
\DeclareRobustCommand{\MakeUppercase}[1]{{%
      \def\i{I}\def\j{J}%
      \def\reserved@a##1##2{\let##1##2\reserved@a}%
      \expandafter\reserved@a\@uclclist\reserved@b{\reserved@b\@gobble}%
      \protected@edef\reserved@a{\uppercase{#1}}%
      \reserved@a
   }}
\DeclareRobustCommand{\MakeLowercase}[1]{{%
      \def\reserved@a##1##2{\let##2##1\reserved@a}%
      \expandafter\reserved@a\@uclclist\reserved@b{\reserved@b\@gobble}%
      \protected@edef\reserved@a{\lowercase{#1}}%
      \reserved@a
   }}
\makeatother
\expandafter\let\csname ver@natbib.sty\endcsname\relax
\usepackage{csquotes}
\usepackage[%
        style=phys,eprint=false,%
        articletitle=true,biblabel=brackets,%
        chaptertitle=true,pageranges=false,%
        sortcites,sorting=none
        ]{biblatex}
\DeclareFieldFormat[article]{title}{\mkbibemph{#1}}
\usepackage{hyperref}
\usepackage[capitalise]{cleveref}
\crefmultiformat{figure}{Figs.~#2#1\xdef\mycreffirstarg{#1}#3}%
 { and~#2{\crefstripprefix{\mycreffirstarg}{#1}}#3}{, #2#1#3}{ and~#2#1#3}
\PassOptionsToPackage{demo}{graphicx}

\usepackage{mathabx}

\def\d{\partial}
\def\nl{\nonumber\\}

\usepackage{framed}
\colorlet{shadecolor}{blue!20}
\usepackage[normalem]{ulem}

\makeatletter
\newcommand\colorwave[1][blue]{\bgroup \markoverwith{\lower3.5\p@\hbox{\sixly \textcolor{#1}{\char58}}}\ULon}
\font\sixly=lasy6 
\makeatother
\makeatletter
\def\@fpheader{\relax}
\makeatother

\makeatletter
\def\CT@@do@color{%
  \global\let\CT@do@color\relax
    \@tempdima\wd\z@
    \advance\@tempdima\@tempdimb
    \advance\@tempdima\@tempdimc
    \advance\@tempdimb\tabcolsep
    \advance\@tempdimc\tabcolsep
    \advance\@tempdima2\tabcolsep
    \kern-\@tempdimb
    \leaders\vrule
    \hskip\@tempdima\@plus  1fill
    \kern-\@tempdimc
    \hskip-\wd\z@ \@plus -1fill }
\makeatother

\hypersetup{%
  colorlinks=true, linktocpage=true, pdfstartpage=1, pdfstartview=FitV,%
  breaklinks=true, pdfpagemode=UseNone, pageanchor=true, pdfpagemode=UseOutlines,%
  plainpages=false, bookmarksnumbered, bookmarksopen=true, bookmarksopenlevel=1,%
  hypertexnames=true, pdfhighlight=/O,
  urlcolor=Cerulean, linkcolor=Cerulean, citecolor=Cerulean, 
}

\begin{document}
\renewcommand{\bibliography}[1]{}
\title{Implications of dark sector mixing on leptophilic scalar dark matter}
\author{Sreemanti Chakraborti}
\email{sreemanti@iitg.ac.in}
\affiliation{Department of Physics,
               Indian Institute of Technology Guwahati,
               Assam, India}
\author{Rashidul Islam}
\email{rislam@iitg.ac.in}
\affiliation{Department of Physics,
               Indian Institute of Technology Guwahati,
               Assam, India}
\affiliation{Department of Physics,
               Mathabhanga College, Cooch Behar,
               West Bengal, India}
%

%
\begin{abstract}
We propose a new viable outlook to the mixing between a singlet and a doublet leptonic dark sector fields. This choice relaxes the dark matter (DM) search constraints on the quintessential scalar singlet DM as well as presents new opportunities for its detection in the lab. The mixing produces an arbitrary mass difference between the two components of the extra doublet in a gauge-invariant way, without introducing any new scale of electroweak symmetry breaking in the theory. It also provides a useful handle to distinguish between the dark sector particles of different isospins, which is a challenging task otherwise. As the dark leptons coannihilate non-trivially, the mixing effectively enhances the viable parameter space for the relic density constraint. In low DM mass regime, our analysis shows that with a non-zero mixing, it is possible to relax the existing indirect search bounds on the upper limit of the DM-Standard Model coupling. From the analysis of the \(3\tau + E^{miss}_T\) and \(\ell\,\tau + E^{miss}_T\) channels for LHC at $\sqrt{s} = 13$~TeV, we show that one ensures the presence of the mixing parameter between the dark sector particles of the theory by looking at the peak and tail positions of the kinematic distributions. Even with a tweak in the values of other free parameters within the viable parameter region, the distinct peak and tail positions of the kinematic distributions remains a constant feature of the model. While both the channels present us the opportunity to detect the mixing signature at the LHC/HL-LHC, the former gives better results in terms of a larger region of mixing parameter. From the fiducial cross section, the projected statistical significance for the integrated luminosity, ${\mathscr L} = 3~\text{ab}^{-1}$, are shown for a combined parameter region obeying all the existing constraints, where there is the best possibility to detect such a signature.
\end{abstract}

\maketitle

%
\section{Introduction}
\label{sec:intro}
The dark matter (DM) constituting 27\% of the energy budget of the Universe is now a settled fact. The cosmological considerations and astrophysical observations have put this matter beyond any doubt. The precisely measured value of the cosmological relic abundance by WMAP \cite{Hinshaw:2012aka} and Planck \cite{Aghanim:2018eyx} is $\Omega_{\rm DM} h^2 = 0.1199 \pm 0.0027$, $h$ being the reduced Hubble constant. The search for a suitable DM candidate is still on as ever \cite{Bergstrom:2000pn, Bertone:2004pz, Feng:2010gw}, and the most widely explored among all is the Weakly Interacting Massive Particle (WIMP) whereas the scalar singlet DM or scalar ``Higgs-portal'' scenario tops among WIMP paradigm \cite{Silveira:1985rk, McDonald:1993ex, Burgess:2000yq}. It is also widely known that the parameter space of the Higgs-portal models has shrunk over the years \cite{Athron:2017kgt, Arcadi:2019lka} by stringent constraints from the direct detection (DD) \cite{Agnes:2015ftt, Akerib:2016vxi, Aprile:2017iyp, Cui:2017nnn}, indirect detection (ID) \cite{Abramowski:2013ax, Ackermann:2015lka, Fermi-LAT:2016uux} and invisible Higgs decay \cite{Belanger:2013xza, Aad:2015pla, Khachatryan:2016whc} searches.

Nonetheless, the Higgs-portal scenario is not entirely out of favour as a model. There are alternatives for evading the existing constraints, namely, by considering additional symmetries \cite{Belanger:2012zr, Bhattacharya:2013hva, Bian:2013wna} or adding new particles \cite{Batell:2016ove, Bandyopadhyay:2017tlq, Han:2018bni} that give rise to new portals for DM annihilation. However, the possibility that we shall address in the present article is the so-called {\em coannihilation} \cite{Griest:1990kh, Ellis:1998kh, Jittoh:2005pq, Kaneko:2008re, Jittoh:2010wh, Citron:2012fg, Konishi:2013gda, Desai:2014uha, Bhattacharya:2017fid, Baker:2018uox, Arganda:2018hdn, Godbole:2008it} mechanism, the one where the DM annihilates with another dark sector particle, and the chemical equilibrium between the annihilating particles ensures the substantial depletion of DM number density. Previously, we have shown how we can successfully evade the existing stringent constraints on scalar singlet DM by introducing a vector-like dark lepton doublet \cite{Chakraborti:2019fnz}. This simple addition to the scalar singlet DM extension of the SM opens up new vistas of possibilities, namely (i) new dark leptons interacting with the SM via gauge interactions introduces new annihilation channels, and (ii) novel Yukawa structure in the Lagrangian enhances cross sections facilitating its search in colliders.

However, the $SU(2)_L$ gauge invariance mandates the degeneracy of the two components of the doublet at the tree level. To lift this degeneracy, in Ref \cite{Chakraborti:2019fnz}, we introduced a $\mathbb{Z}_2$-even scalar triplet. The said scalar triplet otherwise does not play any role in the DM phenomenology, rendering the exercise ad hoc. Here we propose a better alternative to the earlier case, which is to add a $\mathbb{Z}_2$-odd singlet fermion instead. In principle, the same charge dark fermions can mix among themselves to give physical eigenstates, which in turn, gives rise to the mass splitting between the pair. This mixing gives rise to rich phenomenological implication, unlike the previous case. Hence, this exercise is not limited to generating finite mass splitting between the components of the doublet. In other words, far from being an ad hoc addition to the model, this addition dictates the outcome of the processes through the mixing angle. Here we are interested in the implications of mixing between the charged dark sector fermions as well as the distinguishability of the pure and the mixed leptonic states from the observations.

To put the matter into perspective, we would like to mention that the mixing between a singlet and a doublet in the context of DM studies is nothing new. Previously, several authors used this structure to explain various phenomena. In supersymmetric (SUSY) theories, the bino can mix with the Higgsinos to give rise to a neutralino \cite{Cirelli:2005uq, ArkaniHamed:2006mb, Cheung:2012qy, Badziak:2017the}. Out from the realm of SUSY, this matter has been taken up by other authors to build up minimal DM models to address shrinking parameter space from relic density and direct detection measurements \cite{Cohen:2011ec, Cheung:2013dua, Bhattacharya:2015qpa, Yaguna:2015mva, Calibbi:2015nha, Banerjee:2016hsk}. In addition to the context of DM, this extension has successfully explained small values of neutrino masses and the related phenomenology \cite{Klasen:2016vgl, Bhattacharya:2016lts, Wang:2016lve, Restrepo:2019soi}. See Ref. \cite{Bhattacharya:2018fus}, for a recent review and an exhaustive list of relevant papers. Now, where we differ from all these works is that they took a mixed dark lepton state as a DM candidate, which limits the number of decay channels. Here, the DM candidate is a separate scalar singlet, which phenomenologically is the same as the quintessential minimal extension of the SM. The dark leptons add a new portal for pair annihilation of DM. Besides, there are coannihilation and mediator annihilation channels, and thus it enriches the DM dynamics. The mixing between the dark partners relaxes the existing experimental limits on the scalar singlet DM scenario. It also provides a handle to identify the viable parameter space dominated by dark sector partners of different isospins.

The probe of the DM candidates in the controlled collider environment is always a challenging task. Here we analyse the situation in the case of Large Hadron Collider (LHC). We focus on how the signatures of the mixing parameter from the kinematic distributions of the relevant observables can help us decipher the DM signals from our model. We will see that the addition of an extra singlet does not complicate the search strategies, but opens up new avenues. The mixing parameter gives an extra handle in tuning the kinematic distributions that will be easily accessible at the LHC. Apart from collider searches, we will see how does it affect the indirect search prospects of the DM.

We organised the paper as follows. \cref{sec:model} gives a detailed account of the model. We discuss DM phenomenology, its formalism, and the observations from the relic density in \cref{sec:relic}. \cref{sec:collider} addresses detection prospect of the model in DM search experiments, e.g., hadron collider and indirect searches. Finally, in \cref{sec:conc} we present our conclusions.

%
\section{Model description}
\label{sec:model}
We discussed briefly in the introduction the motivation for the choice of our model content. Here we shall develop from that motivation a detailed description of the model parameters. As mentioned previously, ours is a leptophilic model that can evade all the existing bounds on scalar singlet DM. For that, we consider dark-sector partner(s) which can either be a doublet \cite{Chakraborti:2019fnz} or a singlet fermion \cite{Khoze:2017ixx}. Previously \cite{Chakraborti:2019fnz} we studied the case of a dark lepton doublet partner which coannihilates with the singlet scalar DM candidate as well as acts as a portal, depending on the parameter space, which results in relaxing the constraints considerably. As the gauge invariance does not allow non-degenerate mass states for the doublet partners at the tree level, we had to introduce a new scale of electroweak symmetry breaking. We added a, otherwise redundant, scalar triplet to generate finite mass splitting between them. This mass splitting plays a significant role in determining the DM signatures in the collider. However, the measurement of the $\rho$ parameter \cite{Tanabashi:2018oca} constrains the value of the mass splitting $\leq 10$~GeV. The motivation to make it arbitrary leads us to add a $\mathbb{Z}_2$-odd singlet fermion instead of a $\mathbb{Z}_2$-even scalar triplet in the particle spectrum. We shall see in this section how this singlet fermion plays a vital role in lifting the degeneracy of the doublet states depending on the parameter space.

Moreover, in Ref \cite{Chakraborti:2019fnz}, we had pointed out that none of the observations was useful to distinguish between the two coannihilating particles. Here, in principle, the physical states of the dark leptons can be a mixed state between the same charge component of the doublet and the singlet. Thus it gives us the most general minimal scenario where we can meet all our demands. We want to distinguish between these two dark leptons from the observations of the DM in experimental searches. In this light, we will show later on, that the pair annihilations are more useful in the collider analysis and indirect searches. Whereas in the relic density scenario, the coannihilation channels can better exhibit the mixing effects. With all these motivations clear in our mind, we give below the parameter content of our model.
\begin{table}[!ht]
  \centering
  {\setlength{\tabcolsep}{1em}
  \begin{tabular}{c|c c c|c c c}
    \hline
              & $\ell_L$ & $e_R$ &  $H$  & $\varPsi$ & $\xi$ & $\phi$       \\
    \hline\hline
    $SU(2)_L$ & $\mathbf{2}$
                         & $\mathbf{1}$
                                 & $\mathbf{2}$
                                         & $\mathbf{2}$
                                                     & $\mathbf{1}$
                                                             & $\mathbf{1}$ \\
    \hline
    $U(1)_Y$  &  $-1/2$  & $-1$  & $1/2$ &  $-1/2$   & $-1$  & $0$          \\
    \hline
    $\mathbb{Z}_2$
              &   $+$    &  $+$  &  $+$  &   $-$     &  $-$  & $-$          \\
    \hline
  \end{tabular}
  }
  \caption{Quantum number assignment of the relevant fields in our model. Electromagnetic charges are given by \(Q = t^3 + Y\).}
  \label{tab:quant}
\end{table}

\cref{tab:quant} shows the particle content and the quantum number assignments of our model. The $\mathbb{Z}_2$-odd dark sector contains a vector-like Dirac fermionic doublet, \(\varPsi^T = (\psi^0, \psi_1)\), a fermionic singlet \(\xi\) and a real scalar singlet \(\phi\). \(\phi\) is our DM candidate which interacts with the SM Higgs via quartic as well as portal coupling and the \(\mathbb{Z}_2\) symmetry renders stability to it. The fields $\varPsi$ and $\xi$ couples with the SM doublets $\ell_L, H$ and the scalar singlet $\phi$ via three different kinds of gauge-invariant Yukawa interactions.

Hence the resulting Lagrangian takes the form
\begin{align}
 {\cal L} &= {\cal L}_{\rm SM}
 + \widebar\varPsi\,(i\slashed D - M_\varPsi)\,\varPsi
 + \widebar\xi\,(i\slashed\d - m_\xi)\,\xi
 \nl
 &
 + \frac{1}{2}\,(\d_\mu \phi)^2 - \frac{\mu^2_\phi}{2} \phi^2 - \frac{\lambda_\phi}{4} \phi^4
 - \frac{\lambda_{h\phi}}{2} (H^\dag H)\,\phi^2
 \nl
 &
 - \left[
   y^D_j\,(\widebar\ell_{j L}\,\varPsi)\,\phi
 + y^S_j\,(\bar e_{j R}\,\xi)\,\phi
 + y\,(\widebar\varPsi\,H)\,\xi
 + {\rm h.c} \right] \,,
 \label{eq:Lag_tot}
\end{align}
where ${\cal L}_{\rm SM}$ is the SM Lagrangian and a sum over the generation index $j$ is implied. \(M_\varPsi = m_{\psi^0} = m_{\psi_1}\), is the degenerate bare mass term of the doublet and \(D_\mu = \d_\mu + ig_W\,t^a\,W^a + ig^\prime\,Y\,B_\mu\) is the covariant derivative. The mass of the scalar singlet $\phi$ is given by
\begin{gather}
  m^2_\phi = \mu^2_\phi + \lambda_{h\phi}\,\frac{v^2}{2}\,.
  \label{eq:dm_mass}
\end{gather}

After electroweak symmetry breaking in Lagrangian \eqref{eq:Lag_tot}, the Yukawa interaction involving the SM Higgs boson induces non-diagonal mass terms between the singlet and the same charge component of the doublet. As a result, the bare mass terms of the said fermionic degrees of freedom are augmented with the mixed terms as shown below.
\begin{align}
  {\cal L}_{\rm mass}
  =
  & m_{\psi_1}\,\widebar\psi_1\psi_1
  + m_\xi\,\widebar\xi\xi
  + \frac{y\,v}{\sqrt{2}}\widebar\psi_1\xi
  + \frac{y\,v}{\sqrt{2}}\widebar\xi\psi_1
\end{align}
The resulting mass matrix in the bare basis $(\psi_1, \xi)$ is given by
\begin{align}
  M_{(\psi_1, \xi)}
  =&
  \begin{pmatrix}
    m_{\psi_1} & \frac{y\,v}{\sqrt{2}} \\
    \frac{y\,v}{\sqrt{2}} & m_\xi
  \end{pmatrix}
\end{align}
To diagonalise the above mass matrix, we introduce the physical basis $(\psi, \chi)$ through the orthogonal transformation
\begin{gather}
\begin{gathered}
  \psi = c_\alpha\,\psi_1 + s_\alpha\,\xi\,,
  \\
  \chi = - s_\alpha\,\psi_1 + c_\alpha\,\xi\,;
\end{gathered}
  \label{eq:mass_basis}
\end{gather}
where $\alpha$ is the mixing angle. Thus, the diagonal mass matrix of the physical mass terms turns out to be
\begin{align}
  M_{(\psi, \chi)}
  =&
  \begin{pmatrix}
    m_\psi & 0 \\
    0 & m_\chi
  \end{pmatrix}
  =
  \begin{pmatrix}
    c_\alpha & s_\alpha \\
    - s_\alpha & c_\alpha
  \end{pmatrix}
  \begin{pmatrix}
    m_{\psi_1} & \frac{y\,v}{\sqrt{2}} \\
    \frac{y\,v}{\sqrt{2}} & m_\xi
  \end{pmatrix}
  \begin{pmatrix}
    c_\alpha & - s_\alpha \\
    s_\alpha & c_\alpha
  \end{pmatrix}\,.
\end{align}
From the above, we get the physical masses \(m_\psi,m_\chi\) and the mixing angle \(\alpha\) in terms of the bare parameters
\begin{gather}
  t_{2\alpha} = \frac{\sqrt{2}\,y\,v}{m_{\psi_1} - m_\xi}\,,
  \label{eq:mixing}
  \\
\begin{aligned}
  m_\psi =& \frac{1}{2} (m_{\psi_1} + m_\xi) + \frac{1}{2} \sqrt{(m_{\psi_1} - m_\xi)^2 + 2\,y^2\,v^2}\,,
  \\
  m_\chi =& \frac{1}{2} (m_{\psi_1} + m_\xi) - \frac{1}{2} \sqrt{(m_{\psi_1} - m_\xi)^2 + 2\,y^2\,v^2}\,.
\end{aligned}
  \label{eq:phys_mass}
\end{gather}
The Yukawa coupling $y$ and the mass $m_{\psi^0}$ are dependent on the above free parameters of the model. On solving \cref{eq:mixing,eq:phys_mass} we get the expressions of these dependent parameters of our model as follows
\begin{gather}
  y = \frac{(m_\psi - m_\chi)\,s_{2\alpha}}{\sqrt{2}\,v}\,,
  \label{eq:y2}
  \\
\begin{aligned}
  m_{\psi_1} =& m_\psi\,c^2_\alpha + m_\chi\,s^2_\alpha = m_{\psi^0}\,,
  \\
  m_{\xi} =& m_\psi\,s^2_\alpha + m_\chi\,c^2_\alpha\,,
\end{aligned}
  \label{eq:unphys_mass}
\end{gather}
in terms of free parameters $s_\alpha, m_\psi$ and $m_\chi$.

As a result of the new interactions from the dark sector in the Lagrangian \eqref{eq:Lag_tot}, we get the new gauge mediated Feynman vertices as follows
\begin{alignat}{4}
\begin{aligned}
  & A_\mu\psi^+\psi^- \mkern-24mu&&: - i e \gamma_\mu,
  &&W^+_\mu\chi^-\widebar\psi^0 \mkern-24mu&&: - \frac{i e s_\alpha}{\sqrt{2} s_W} \gamma_\mu,
  \\
  & A_\mu\chi^+\chi^- \mkern-24mu&&: - i e \gamma_\mu,
  &&W^+_\mu\psi^-\widebar\psi^0 \mkern-24mu&&: \frac{i e c_\alpha}{\sqrt{2} s_W} \gamma_\mu,
  \\
  & Z_\mu\psi^0\widebar\psi^0 \mkern-24mu&&: \frac{i e}{s_{2W}} \gamma_\mu,
  &&Z_\mu\psi^+\psi^- \mkern-24mu&&: - \frac{i e (s^2_\alpha - c_{2W})}{s_{2W}} \gamma_\mu,
  \\
  & Z_\mu\psi^+\chi^- \mkern-24mu&&: \frac{i e c_\alpha s_\alpha}{s_{2W}} \gamma_\mu,
  &&Z_\mu\chi^+\chi^- \mkern-24mu&&: \frac{i e (c^2_\alpha - c_{2W})}{s_{2W}} \gamma_\mu.
\end{aligned}
\label{FR_Gauge}
\end{alignat}
And vertices coming from the new Yukawa interactions that will play a significant role in our analysis are
\begin{alignat}{4}
\begin{aligned}
  & \phi\tau^+\psi^- \mkern-24mu&&: - i ( c_\alpha y^D_\tau P_R + s_\alpha y^S_\tau P_L ),
  \mkern-6mu&&h\psi^+\psi^- \mkern-24mu&&: - \frac{i y s_{2\alpha}}{\sqrt{2}},
  \\
  & \phi\tau^+\chi^- \mkern-24mu&&: - i ( - s_\alpha y^D_\tau P_R + c_\alpha y^S_\tau P_L ),
  \mkern-6mu&&h\psi^+\chi^- \mkern-24mu&&: - \frac{i y c_{2\alpha}}{\sqrt{2}},
  \\
  & \phi\nu_\tau\widebar\psi^0 \mkern-24mu&&: - i y^D_\tau P_L,
  \mkern-6mu&&h\chi^+\chi^- \mkern-24mu&&: \frac{i y s_{2\alpha}}{\sqrt{2}}.
\end{aligned}
\label{FR_Higgs}
\end{alignat}

Apart from the those given in \cref{eq:dm_mass,eq:mixing,eq:phys_mass}, we have only two more free parameters in our model, namely, the third generation Yukawa couplings $y^{D,S}_\tau$. We put the Yukawa couplings of light leptons at $y^{D,S}_e \sim y^{D,S}_{\mu} \lesssim 10^{-9}$ to conform the muon $g-2$ measurements \cite{Tanabashi:2018oca}. All the Yukawa couplings are to remain in the perturbative regime such that $y^{D,S}_\tau, y \leq 4\pi$. A conservative choice of $\lambda_{h\phi} \lesssim 10^{-4}$ is in place all through our analysis to keep the bounds from the direct detection searches and the invisible decay measurements at bay.

%
\section{Relic density analysis}
\label{sec:relic}

Due to the presence of more than one dark sector particles, the DM number changing processes, in this model, are three-fold: (i) pair annihilation ($\phi \phi \to \rm SM\,SM$), (ii) coannihilation ($\phi \psi^{\pm0} \to \rm SM\,SM$), and (iii) mediator annihilation ($\psi^{\pm0} \psi^{\mp0} \to \rm SM\,SM$). Keeping in mind the assumption of thermal freeze-out, that the dark sector particles are in equilibrium with the thermal bath in the early Universe, whereas, in chemical equilibrium with each other, one can write the Boltzmann equation describing the number density, $n$ of the DM as follows~\cite{Griest:1990kh}.
\begin{gather}
  \frac{dn}{dt}
  =
  - 3\,H\,n - \ev{\sigma_{\rm eff}\,v}\,(n^2-n^2_{\rm eq}) \,.
\end{gather}
The effective velocity-averaged annihilation cross section, $\ev{\sigma_{\rm eff}\,v}$ can be written as
\begin{align}
  \ev{\sigma_{\rm eff}\,v}
  =
  \Big(\sum\limits_i\widebar{g}_i\Big)^{-2}\,\frac{1}{2}\,\sum\limits_{i,j}\,\widebar{g}_i\,\widebar{g}_j\,\ev{\sigma_{ij\to\rm SM\,SM}} \,,
\label{BEQ}
\end{align}
where the indices $i,j$ denote any of the dark sector particles $\phi$, $\psi^{\pm0}$ or $\chi^\pm$, and
\begin{gather}\label{g_BEQ}
\begin{gathered}
  \widebar g_i = g_i\,(1 + \delta m_i/m_\phi)^{3/2}\,\exp[-x\,\delta m_i/m_\phi] \,,
  \\
  \delta m_i = (m_i - m_\phi)\,,\qquad x=m_\phi/T \,,
  \\
  g_\phi=1\,,\qquad g_{\psi^\pm}=g_{\chi^\pm}=g_{\psi_0}=2 \,.
\end{gathered}
\end{gather}
For our analyses, we have implemented the Lagrangian \eqref{eq:Lag_tot} along with all the above relations (\cref{eq:dm_mass,eq:mass_basis,eq:mixing,eq:phys_mass,eq:y2,eq:unphys_mass}) in \textsc{FeynRules}~\cite{Alloul:2013bka}. Using the resulting model file, we carried out all the following DM analysis with the help of \textsc{micrOMEGAs}~\cite{Belanger:2014vza}.

\begin{figure}[!ht]
 \centering
 \includegraphics[height=0.2\textwidth]{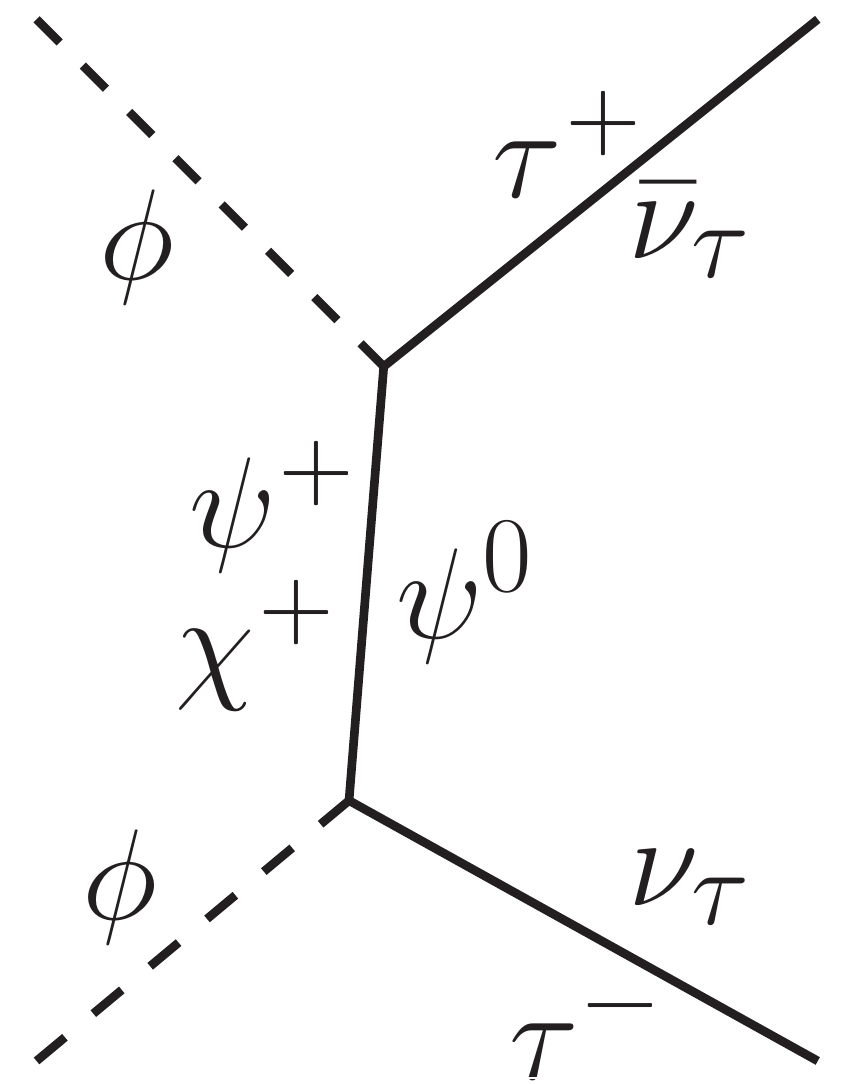}
 \caption{Pair annihilation possibilities}
 \label{diag:ann}
\end{figure}

As discussed in the previous section, mixing affects only the charged dark fermions. Therefore $\ev{\sigma_{\rm eff}\,v}$ will be sensitive to mixing for the annihilation channels which involve these dark fermions in the initial state and/or the propagator. One such possibility is the pair annihilation of $\phi$, where the charged dark fermions appear in the $t$-channel propagator~(\cref{diag:ann}). 

\begin{figure}[!ht]
  \centering
  \subfloat[]{\includegraphics[height=0.18\textwidth]{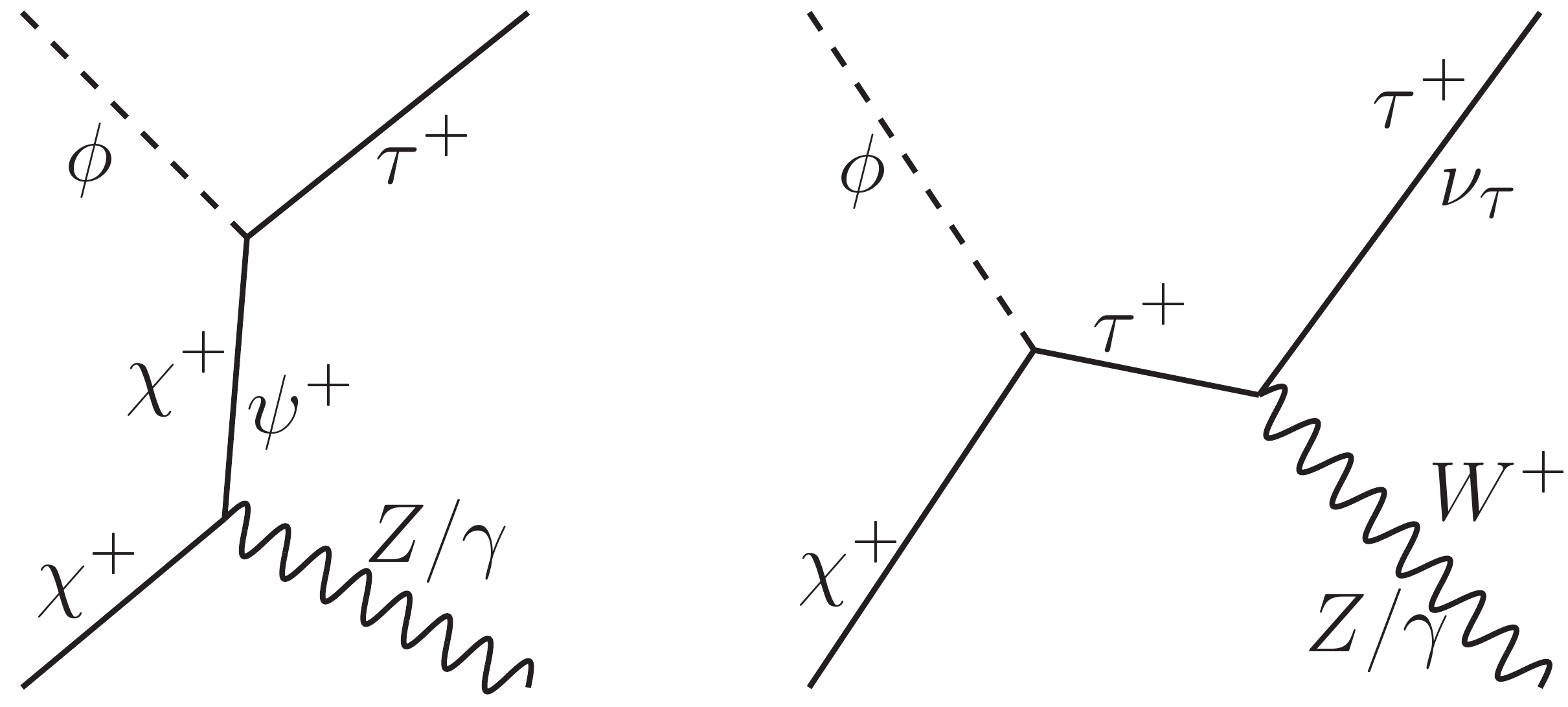}\label{diag:coann3}}
  \hskip1em
  \subfloat[]{\includegraphics[height=0.18\textwidth]{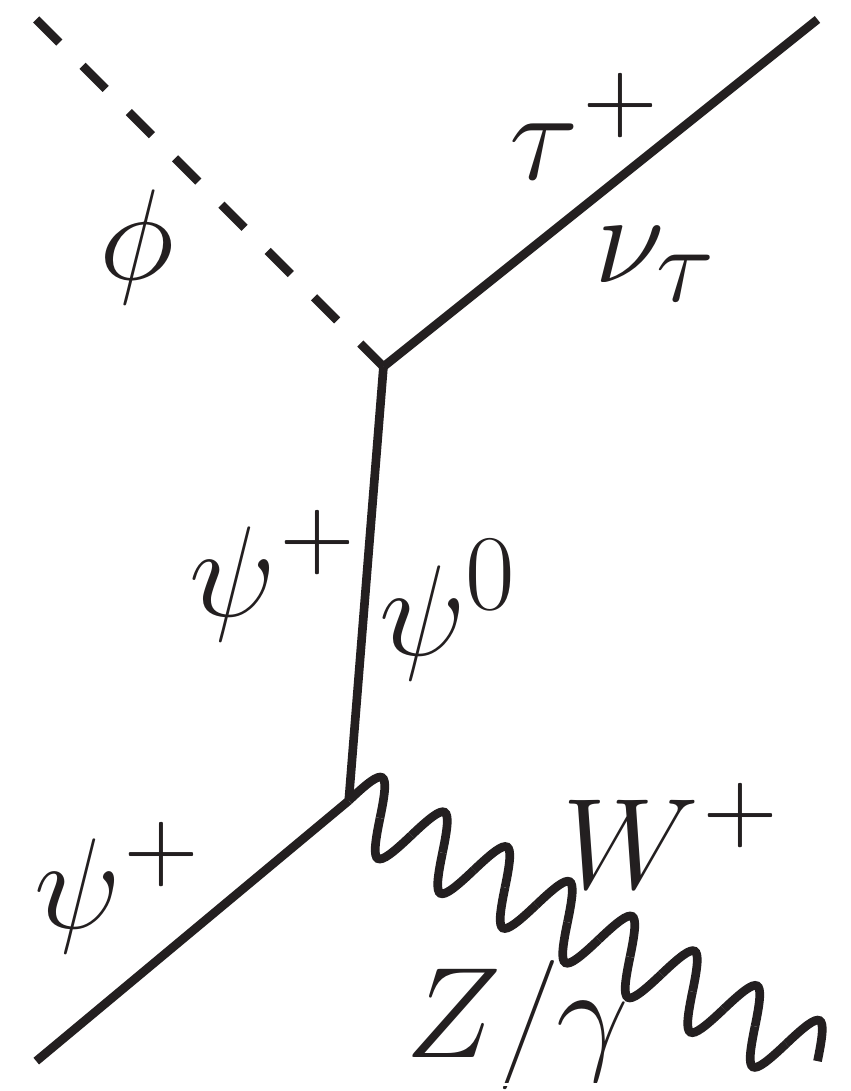}\label{diag:coann1}}
  \hskip1em
  \subfloat[]{\includegraphics[height=0.18\textwidth]{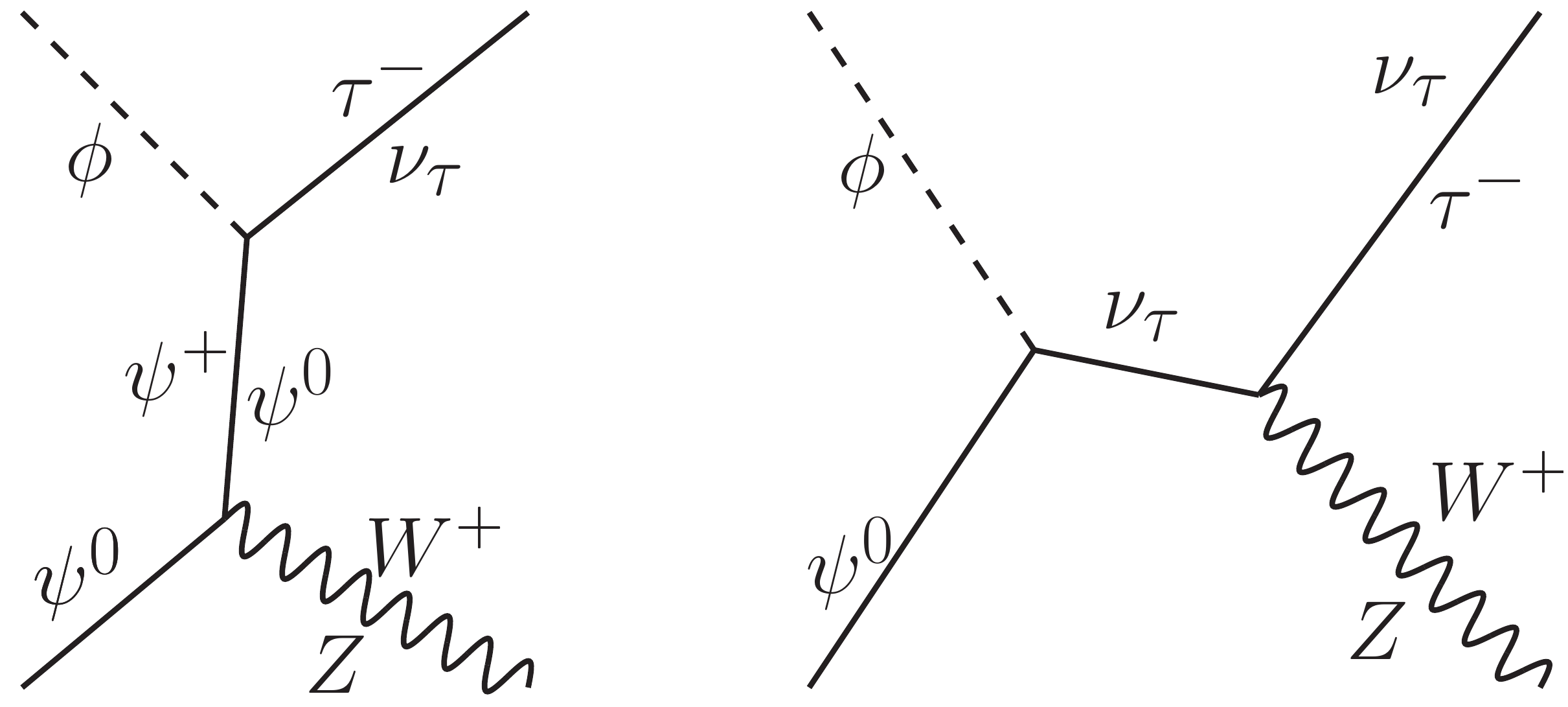}\label{diag:coann2}}
  \caption{ Coannihilation channels that are non-trivially dependent on mixing. \cref{diag:coann3,diag:coann1} depict DM-charged fermion coannihilation whereas \cref{diag:coann2} represents DM-neutral fermion interaction.}
  \label{diag:coann}
\end{figure}

But the mixing effects are most prominent if one considers the coannihilation channels. This is because in some of the coannihilation diagrams~(\cref{diag:coann3,diag:coann1}), in addition to the propagators, one of the initial state particles are directly affected by mixing. Therefore, the coannihilation channels of $\phi$ with $\psi^+$ and $\chi^+$ amply show the mixing effects.\footnote{Throughout this article, we denote the charged dark leptons by $+$ sign. However, All the arguments are equally applicable for negatively charged particles as well.} On the other hand, $\psi^0$ not being a mixed state, $\phi\, \psi^0$ coannihilation channel~(\cref{diag:coann2}) is much less affected by mixing. As we are interested to study the mixing effects on the phenomenology, we will mostly concentrate on the coannihilation channels for our DM analysis. This implies that the mass splitting between DM and the dark fermions should be small throughout the study ($\lesssim 30$ GeV) and the dark sector-SM coupling is not very large ($\lesssim 1$).

\begin{figure}[!ht]
  \centering
  \includegraphics[width=0.8\textwidth]{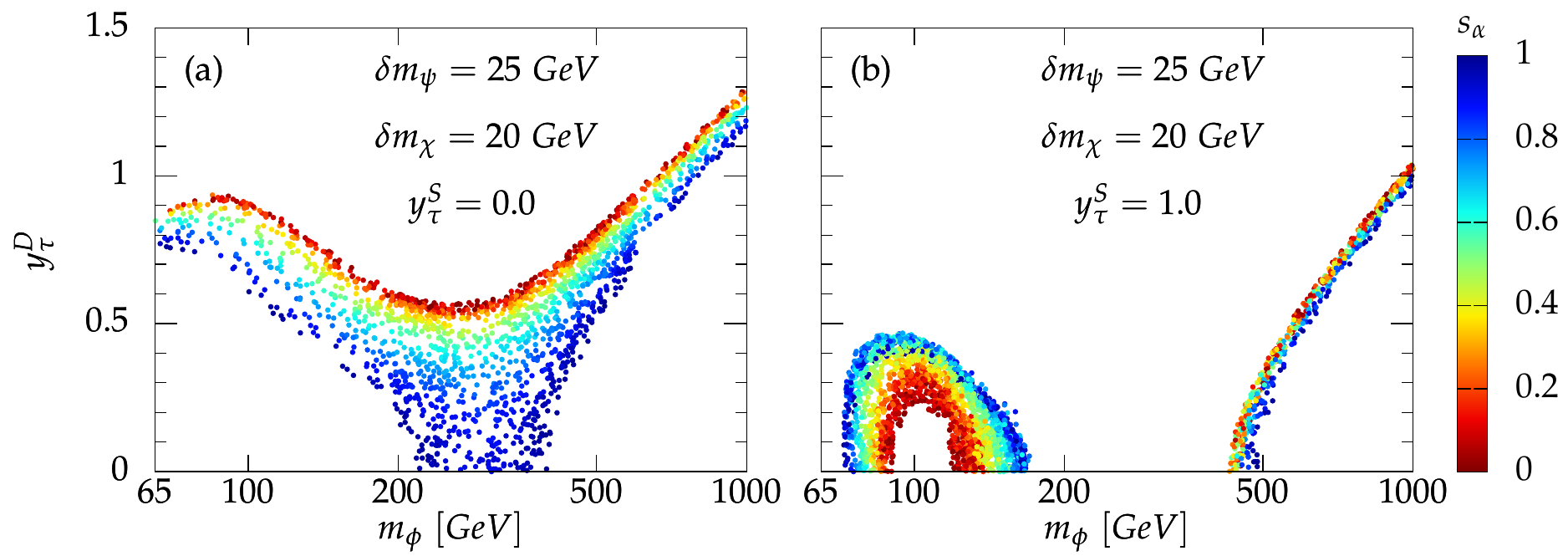}
  \caption{$y_\tau^D$ vs $m_\phi$ correlation for relic density allowed points. In \cref{coann_plot_1}a), $\phi \psi^+$ coannihilation survives for zero mixing and $\phi \chi^+$ coannihilation dominates for non-zero values of $s_\alpha$. In \cref{coann_plot_1}b), due to large $y_\tau^S$, pair annihilation contribution becomes prominent for the large $m_{\phi}$.}
  \label{coann_plot_1}
\end{figure}

For fixed mass splittings between the dark sector particles, the DM-SM Yukawa coupling vs DM mass correlation is discussed amply in the literature \cite{Khoze:2017ixx,Bhattacharya:2017fid}. In Ref. \cite{Chakraborti:2019fnz}, the transition between the three kinds (pair, co- and mediator annihilation) of DM annihilation  is depicted for a similar model with only a lepton doublet as the coannihilating partner of the scalar DM. In this work, this dynamics becomes more interesting in the presence of the charged singlet and doublet dark fermion mixing. In \cref{coann_plot_1}, $y_\tau^D$ is plotted against $m_\phi$ for fixed $\delta m$ and $y_\tau^S$. In the absence of mixing, depicted by the red line in \cref{coann_plot_1}a), the scenario is identical to Ref. \parencite[Fig. 4]{Chakraborti:2019fnz}. In this case, apart from the subdominant pair annihilation, only $\phi \psi^+$ coannihilation contributes to relic density, $\phi \chi^+$ channel being redundant due to $y_\tau^S$=0. However, as mixing increases, both $\phi \psi^+$ and $\phi \chi^+$ coannihilation channels contribute. For our choice of parameters, $\delta m_{\chi} \le \delta m_{\psi}$, hence $\phi \chi^+$ coannihilation (\cref{diag:coann3}) is more dominant than the $\phi \psi^+$ counterpart for a non-zero mixing. This is because both $\phi\psi^+$ and $\phi \chi^+$ coannihilations are predominantly controlled by the channels which have $W^+$ in the final state. Now, for $\phi \psi^+$ coannihilation case, these channel exists even for $s_\alpha = 0$, whereas, for $\phi \chi^+$ case, this channel is realised largely through mixing. Therefore, the mixing effect is more prominent in $\phi \chi^+$ coannihilation scenario over the $\phi \psi^+$ case.

With $y_{\tau}^S$ set to zero, as one goes for larger mixing angle, $\phi \chi^+ \tau^-$ coupling (\cref{FR_Higgs}) increases. This, along with the fact that both dark fermions now can non-trivially coannihilate, the total DM annihilation cross section effectively increases. Hence, for a fixed $m_\phi$, larger mixing corresponds to smaller coupling to be relic density allowed. Therefore, $y_\tau^D$ gradually decreases with increase in $s_\alpha$.

The scenario changes substantially if the same correlation is drawn with a fixed non-zero $y_\tau^S$. In \cref{coann_plot_1}b), we assign a large value to $y_\tau^S$. For such a large coupling, pair annihilation has a substantial contribution to the relic density. This is due to the fact that in this model, in the expression for annihilation cross section, the functional dependence on the DM-SM Yukawa coupling ($\lambda$, let us suppose) is $\lambda^4$ in pair annihilation, while for coannihilation, it is $\lambda^2$. This is also in agreement with \parencite[p.5, p.6]{Chakraborti:2019fnz} where we showed that for a fixed $\delta m$ and variable $m_{\phi}$, coannihilation is dominant for small $m_{\phi}$ and at large DM mass, pair annihilation takes over. In this plot, as $y_\tau^S$ is already large, $y_\tau^D$ is not required to be as high as that in \cref{coann_plot_1}a). Now, for small $m_{\phi}$ values, $\phi \chi^+$ has the dominant contribution in relic density. $\phi \chi^+ \tau^-$ coupling (\cref{FR_Higgs}) is the largest for $s_\alpha$ = 0 and due to this, in order to achieve right relic, the required value of $y_\tau^D$ is the smallest, as we see in this plot for the red band around $m_{\phi}\le$ 180 GeV. As the mixing increases, the above coupling decreases, therefore larger $y_\tau^D$ is required. However, as $m_{\phi}$ increases further, pair annihilation takes over and the mixing effect is lost in the ``tail'' region of the correlation around $m_{\phi} >$ 500 GeV. As discussed above, this is because mixing does not play a significant role in the pair annihilation channels.

\begin{figure}[!ht]
  \centering
  \includegraphics[width=0.8\textwidth]{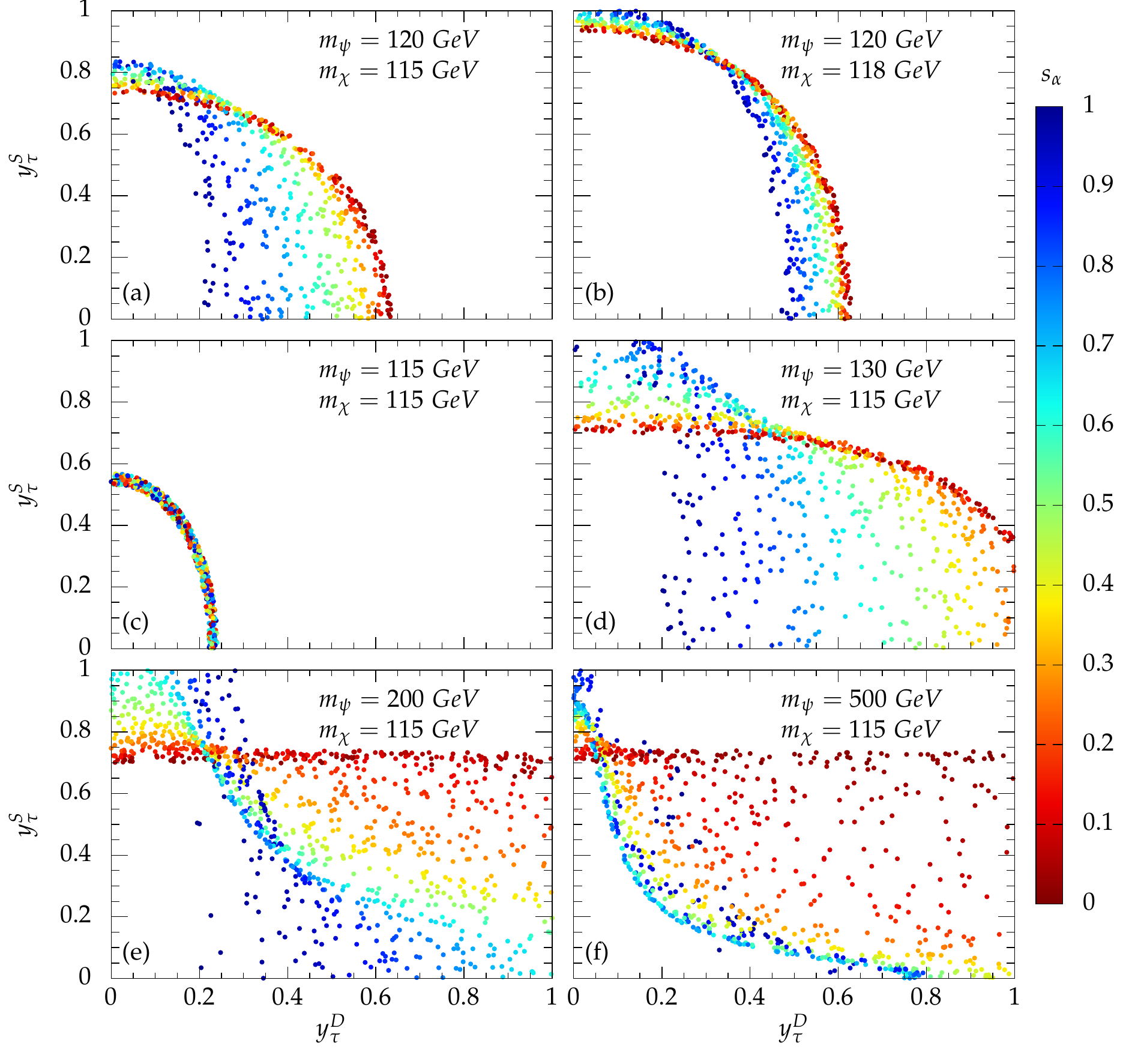}
  \caption{$y_\tau^D$ vs $y_\tau^S$ correlation for relic density allowed points. For fixed DM mass, the variation of the charged dark fermion masses and mixing angle substantially changes the correlation. $m_\phi$ is fixed at 100~GeV.}
  \label{coann_plot_3}
\end{figure}

In \cref{coann_plot_1}, we considered constant mass splittings between the dark sector particles. Now let us discuss how different values of these mass splittings affect the DM dynamics through mixing. In \cref{coann_plot_3}, $y_\tau^D$ and $y_\tau^S$ correlation is plotted as a function of $s_\alpha$ and for different $\delta m$'s. Similar to \cref{coann_plot_1}, here also we have chosen $\delta m_\chi \le \delta m_\psi$ to facilitate the $\phi\,\chi^+$ coannihilation. As already discussed, large mass splitting between DM and dark fermions will suppress the coannihilation contribution towards relic density and larger coupling is required to compensate for that deficit. Hence, depending on the value of $m_{\chi}$ and $m_{\psi}$, the shift of the relic allowed points along $y_\tau^S$ axis is straightforward. Rather, the distribution of the points along the $y_\tau^D$ axis is an interesting feature.

For low mixing, the dominant $\phi \chi^+$ coannihilation channel is $\phi \chi^+ \to \tau^+Z(\gamma)$ (\cref{diag:coann3}) and its cross section is almost proportional to $(y_\tau^S)^2$. This satisfies relic density even for $y_\tau^D=0$ as one can see from the red band in the subfigures of \cref{coann_plot_3}. As the mixing increases, initially, there is a contest between the two terms in $\phi\,\tau^+ \chi^-$ coupling \cref{FR_Higgs}. For low $y_\tau^D$, as mixing increases, to compensate for the overall reduction of the coupling, we see a slight increase in $y_\tau^S$. But as $y_\tau^D$ increases further, the second term in the expression for the coupling becomes gradually negligible, making the coupling almost proportional to $s_\alpha\ y_\tau^D$ for large mixing. Therefore we observe that to keep within the relic bounds, for large mixing and large $y_\tau^D$, $y_\tau^D$ decreases as $s_\alpha$ gradually increases. It is clear in all the plots except \cref{coann_plot_3}c) that the green and blue points shift to the left along $y^D_\tau$ axis as $s_\alpha$ gradually increases.

Now, we observed that in \cref{coann_plot_3}a), where $m_{\psi}-m_{\chi}$ = 5 GeV, $\phi \chi^+$ coannihilation is the dominant channel. But as this splitting decreases to 2 GeV in \cref{coann_plot_3}b), $\phi \psi^0$ channel also contributes substantially. This is because of the coannihilation channels having $W$ boson in the final state, which are available for $\phi \psi^0$ coannihilation even without mixing but accessible largely through mixing for $\phi \chi^+$ coannihilation. The presence of these channels suppresses the mixing dependence in the $y_\tau^D$ vs $y_\tau^S$ correlation, causing reduction in the spread of multicoloured points in \cref{coann_plot_3}b) compared to \cref{coann_plot_3}a). This spread becomes zero in \cref{coann_plot_3}c), where $m_{\psi}$ and $m_{\chi}$ becomes almost degenerate. As already argued, here $\phi\,\psi^0$ coannihilation is the dominant channel in this case with $W^\pm$ in the final state and this makes the correlation completely independent of mixing.

For \cref{coann_plot_3}d) - f), $\delta m_\psi$ is large so that $\phi\psi^+$ coannihilation contribution is very small. This establishes $\phi\,\chi^+ \to \rm SM\,SM$ as the only active coannihilation channel. In absence of $\phi \psi^+$ (hence also $\phi \psi^0$) coannihilation and  for low mixing, the only surviving $\phi \chi^+$ coannihilation in $\ev{\sigma_{\rm eff}\,v}$ has little or no dependence on $y_\tau^D$. The points corresponding to very low mixing, amply show this effect  as the red band gradually flattens out with increasing $\delta m_{\psi}$ as we go from \cref{coann_plot_3}d) - f). Ultimately in \cref{coann_plot_3}f), the red band widens and becomes parallel to $y_\tau^D$ axis and the blue band corresponding to large mixing becomes the thinnest.

\begin{figure}[!ht]
  \centering
  \includegraphics[width=0.8\textwidth]{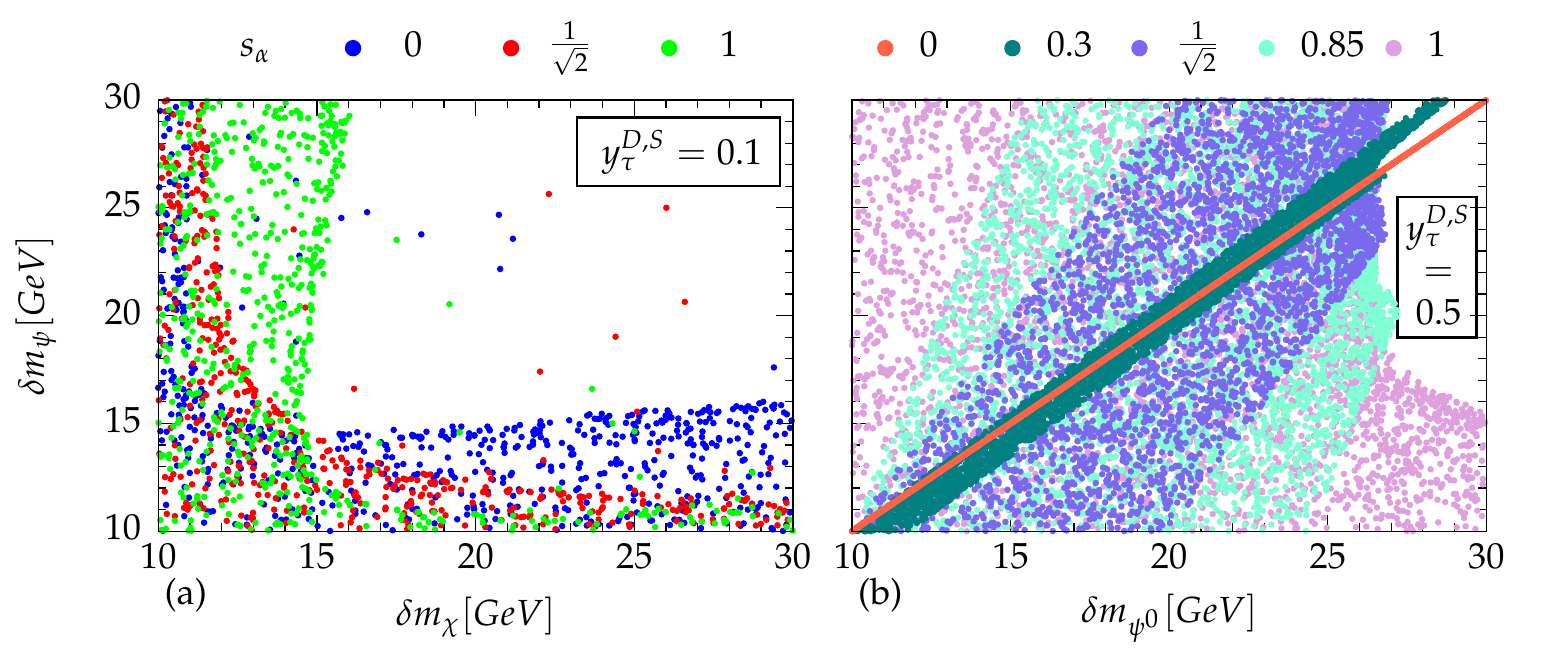}
  \caption{Correlations $\delta m_\chi$ vs $\delta m_\psi$ in \cref{coann_plot_4}a) and $\delta m_{\psi_0}$ vs $\delta m_\psi$ in \cref{coann_plot_4}b) for relic density allowed points as a function of the mixing angle. Since all other parameters are the same for the dark leptons discussed in each plot, it is only up to the mixing parameter to determine the dominant relic density contribution between them. The DM mass varies in the range 65 GeV $\le m_\phi \le$ 1~TeV.}
 \label{coann_plot_4}
\end{figure}

Finally, let us see the effect of variable $\delta m$'s instead of discrete intervals discussed so far. The two plots in \cref{coann_plot_4} show two such correlations of the $\delta m$'s as a measure of mixing. \cref{coann_plot_4}a) focusses on the asymmetry of $\delta m_\chi$ vs $\delta m_\psi$ correlation as a measure of mixing. Mixing plays a vital role here in the distinction between the dark fermions, unlike Ref \cite{Chakraborti:2019fnz}, where such a discrimination was not possible. The two $\delta m$'s vary over the same range and the Yukawa couplings $y_\tau^D$ and $y_\tau^S$ are fixed at the same value. This implies that apart from the mixing parameter, $\chi^+$ and $\psi^+$ stand on equal footing in the context of coannihilation possibilities. However, on fixing different values for the mixing, we see that the plot shows a clear distinction between the two dark leptons of different isospins. In absence of mixing, i.e., for the blue points, the relic density allowed region along $\delta m_\chi$ axis broadens with the increase of $s_\alpha$ whereas along $\delta m_\psi$ it narrows down. This is attributed to the fact that for $s_\alpha = 0$, there are more diagrams in $\phi\psi^+$ coannihilation channel than the $\phi\chi^+$ possibility. These extra diagrams (\cref{diag:coann1}) arise due to Gauge couplings, which are exclusive to $\phi\psi^+ \to$ SM~SM coannihilation because in absence of mixing, $\psi^+$ is purely part of the $SU(2)_L$ doublet and $\chi^+$ is purely a singlet. It is this extra diagram that makes $\phi \psi^+$ coannihilation stronger than the $\phi \chi^+$ counterpart for $s_\alpha = 0$. This can also be verified from the distribution of the blue points, where we see that points at $\delta m_{\chi}$ = 30~GeV correspond to $\delta m_{\psi}$ up to 15 GeV, whereas, points at $\delta m_{\psi}$ = 30~GeV corresponds to $\delta m_{\chi}$ only up to 12 GeV.

We have explicitly checked the above for a few benchmark points and observed that for $s_\alpha = 0$, $\phi \psi^+ \to  W^+ \nu_\tau$ is stronger than any of the $\phi \chi^+ \to$ SM SM channels if the two Yukawa couplings are equal. For $s_\alpha = 1/\sqrt{2}$, on the other hand, $\phi\psi^+$ and $\phi\chi^+$ coannihilation contribution becomes equal even for these extra Gauge channels, so the red points show a symmetric distribution along both the axes. For the green points, i.e., for $s_\alpha = 1$, we simply see the opposite of the $s_\alpha = 0$ case, because $W^+\psi^-\psi^0\big{|}_{s_\alpha = 0}=W^+\chi^-\psi^0\big{|}_{s_\alpha = 1}$. This also agreed with our observation of benchmark point results, that the $\phi\chi^+$ coannihilation is more dominant than $\phi \psi^+$ counterpart for $s_\alpha = 1$. To sum up, we can conclude that the relic density allowed region for the pure leptonic eigenstates widens along $\delta m_{\chi}$ axis, whereas the mixed states tend to widen gradually along $\delta m_{\psi}$ axis.

\cref{coann_plot_4}b) on the other hand, discusses $\delta m_{\psi^0}$ vs $\delta m_{\psi}$ correlation as a function of the mixing parameter. The DM mass varies in the same range as \cref{coann_plot_4}a), and the Yukawa couplings are fixed at $y_\tau^D=y_\tau^S$ = 0.5. 

For $s_\alpha = $ 0, $m_{\psi_0}$ = $m_{\psi}$, as obvious from \cref{eq:phys_mass}. Therefore, the nature of the correlation will be a straight line with a $45^\degree$ slope, which is exactly what the orange line represents. However, for the non-zero $s_\alpha$ values, we see distribution on both sides of the red line. As we can see, the spread of the points varies with $s_\alpha$.

It is to be noted that if the relic density constraint is not imposed, then for a fixed $s_\alpha$, the points would be distributed symmetrically around the orange line, maintaining it as the line of symmetry. This is obvious from the model description arguments of \cref{sec:model}. However, on adding the constraint, one can see from the plot that this symmetry is lost. For large $\delta m$'s, we observe that the relic allowed region shifts slightly to the left of the orange line for all non-zero $s_\alpha$ values.

We numerically checked that for fixed Yukawa couplings, as mixing increases in the range $0 \le s_\alpha \le 1/\sqrt{2}$. the contribution of $\phi \psi^+ \to$ SM SM channels rapidly decreases, but $\phi \psi^0\to$~SM~SM remains much less affected, except a very small increase. Our relic density calculation for a few benchmark points within the allowed region confirmed that $\phi \psi^0 \to W^+ \tau^-$ channel is the dominant channel in $\ev{\sigma_{\rm eff}\,v}$ for this $s_\alpha$ window.

This explains the shift at large $\delta m$ for the non-zero $s_\alpha$ values. As $\phi \psi^+$ contribution decreases, to maintain relic abundance in the observed range, one needs stronger $\phi \psi^0$ coannihilation. This is why the region with a very large $\delta m_{\psi^0}$ remains out of the relic allowed regime, the annihilation cross-section being insufficient, leading to overabundance. It is also justified that we see the upper limit of relic allowed $\delta m_0$ decreasing gradually as $s_\alpha$ increases from 0 to $1/\sqrt{2}$.

On the other hand, we observed that $\phi \chi^+\to$~SM~SM contribution, which was negligible so far, becomes substantial for large mixing, i.e., in the range $1/\sqrt{2} \le s_\alpha \le 1$. As already explained, $\phi \psi^0$ coannihilation remains mostly unaffected by mixing. Therefore, as $\phi \chi^+$ contribution gradually increases with mixing in the above range, it also relaxes the exclusion limit for large $\delta m_{\psi^0}$, which can be seen for the cyan and pink points. Ultimately, for $s_\alpha = 1$, the spread of the relic allowed region is maximum. This is because the relic density is almost entirely dominated by $\phi \chi^+$ coannihilation here, so even large $\delta m_\psi$ and $\delta m_{\psi^0}$ values are allowed except for the white excluded region on the right, where both the $\delta m$'s are very large, leading to insufficient $\ev{\sigma_{\rm eff}\,v}$.

The above features are model-independent and can be generalised for any singlet DM coannihilation scenario that involves a singlet as well a doublet coannihilating partner.

%
\section{Experimental signatures}
\label{sec:collider}
\subsection{DM signatures at the LHC}
\label{sec:lhc}
The problem of finding signatures of DM in a collider environment is a very challenging task. Disentangle the DM signals from a multitude of invisible particles is notoriously hard. Several search strategies are there for this task which addresses the problem from the perspective of the hadron colliders (LHC and future hadron colliders) as well as lepton colliders which are for the very purpose where hadron collider has limitations. In hadron colliders, we do not have access to the longitudinal component of the missing momenta. Hence the task of finding the signature of DM is more difficult as we have to depend entirely on the observables constructed from the transverse momentum components. One can separate {\em multilepton} + {\em missing energy} signals from the backgrounds even for a small signal cross section through advanced techniques of Multivariate Analysis \cite{Chakraborti:2019fnz}. Here we look for specific distributions through which we can decipher the DM signals in a collider environment. We discussed previously that the presence of a doublet along with a singlet dark fermions and the mixing between them plays a significant role in the phenomenology of our model. Here we point to the ways to find out those signatures in collider environments that will highlight this feature of the model and try to distinguish its effect from the distributions. The peaks and end-points of a kinematic distribution can be associated with the masses of the mediating particles. The kinematic distributions of transverse momentum, $p_T$, transverse mass, $m_T$ and invariant mass are a few very significant distributions to study.

In the following we will discuss the relevant distributions in a hadron collider environment like LHC. To perform the analysis for the LHC at the CM energy $\sqrt{s} = 13$~TeV, we proceeded as follows: (1)~\textsc{FeynRules} \cite{Alloul:2013bka} has been used to generate model files. (2)~Events have been generated using \textsc{MadGraph5} \cite{Alwall:2014hca} and showered with \textsc{Pythia 8} \cite{Sjostrand:2014zea}. We used the dynamic factorisation and renormalisation scale for the signal as well as the background events. (3)~The detector simulation has been performed with the help of \textsc{Delphes} \cite{deFavereau:2013fsa}. (4)~The distributions were drawn with the help of \textsc{MadAnalysis 5} \cite{Conte:2012fm}.

In the following analysis of {\em multilepton} + {\em missing energy} signals at the LHC, our focus is particularly on the distributions of the $\tau$'s. As we discussed in \cref{sec:model}, the couplings of the first two generations of the leptons are very constrained by the muon $g - 2$ measurements, whereas the $\tau$ couplings are unbounded. Although phenomenologically interesting in our case, the study of $\tau$ poses many challenges as well. As it is heavier than most of the light quark mesons, $\tau$ can decay hadronically, unlike other leptons. Moreover, the leptonic decays of $\tau$ are difficult to distinguish from prompt leptons in an $\ell~+~E^{miss}_T$ final state. This makes the hadronically decaying $\tau$'s suitable for the collider signatures. We reconstructed the $\tau$-tagged jets from \textsc{Delphes} \cite{deFavereau:2013fsa} with the help of \textsc{FastJet} \cite{Cacciari:2011ma} using the anti-$k_T$ algorithm, with the separation $\Delta R$ of two adjacent $\tau$-jets to be $0.4$. The distance between two objects $i$ and $k$ is defined as $\Delta R_{ik} = \sqrt{(\eta_i - \eta_k)^2 + (\phi_i - \phi_k)^2}$, where $\eta_i$ and $\phi_i$ are the rapidity and azimuthal angle of the object $i$, respectively. In \cref{tab:fid} below, we give the selection criteria, which has been used throughout the following analysis. The $\tau$-tagging efficiency is assumed to be flat and taken to be 60\%, with fake rates from light jets to the $\tau$-jets of 1\%. The statistical significance is given by ${\cal S} = \sqrt{2\left[(S+B)\ln(1+S/B) - S\right]}$, where $S$ and $B$ respectively are the number of expected signal and background events at a particular integrated luminosity ${\mathscr L}$.

\begin{table}[!ht]
  \centering
  {\setlength{\tabcolsep}{1em}
  \begin{tabular}{ c || c | c | c }
    \hhline{-||---}
    Selection parameter & $p^{\ell~(~j~)}_T$&$|\eta_{\ell~(~j~)}|$ & $\Delta R_{ik}$ \\
    \hhline{-||---}
    Cut value           & $10~(~20~)$~GeV   &         $5$          &      $0.4$      \\
    \hhline{-||---}
  \end{tabular}
  }
  \caption{The selection criteria used throughout the analysis, where $\ell = e, \mu$ and $j=u, d, c, s$ and $g$. The distance between two objects $i$ and $k$ is defined as $\Delta R_{ik} = \sqrt{(\eta_i - \eta_k)^2 + (\phi_i - \phi_k)^2}$, where $\eta_i$ and $\phi_i$ are the rapidity and azimuthal angle of the object $i$, respectively.}
  \label{tab:fid}
\end{table}

For the effect of mixing angle $\alpha$ in the collider signatures, two kinds of processes will be useful (1) the processes that are predominantly $W$ boson mediated, and (2) those that are also predominantly $Z$ boson mediated.

\subsubsection{\texorpdfstring{\(3\tau + E^{miss}_T\)}{3 tau + MET} channel}
\begin{figure}[!ht]
 \centering
 \includegraphics[height=0.2\textwidth]{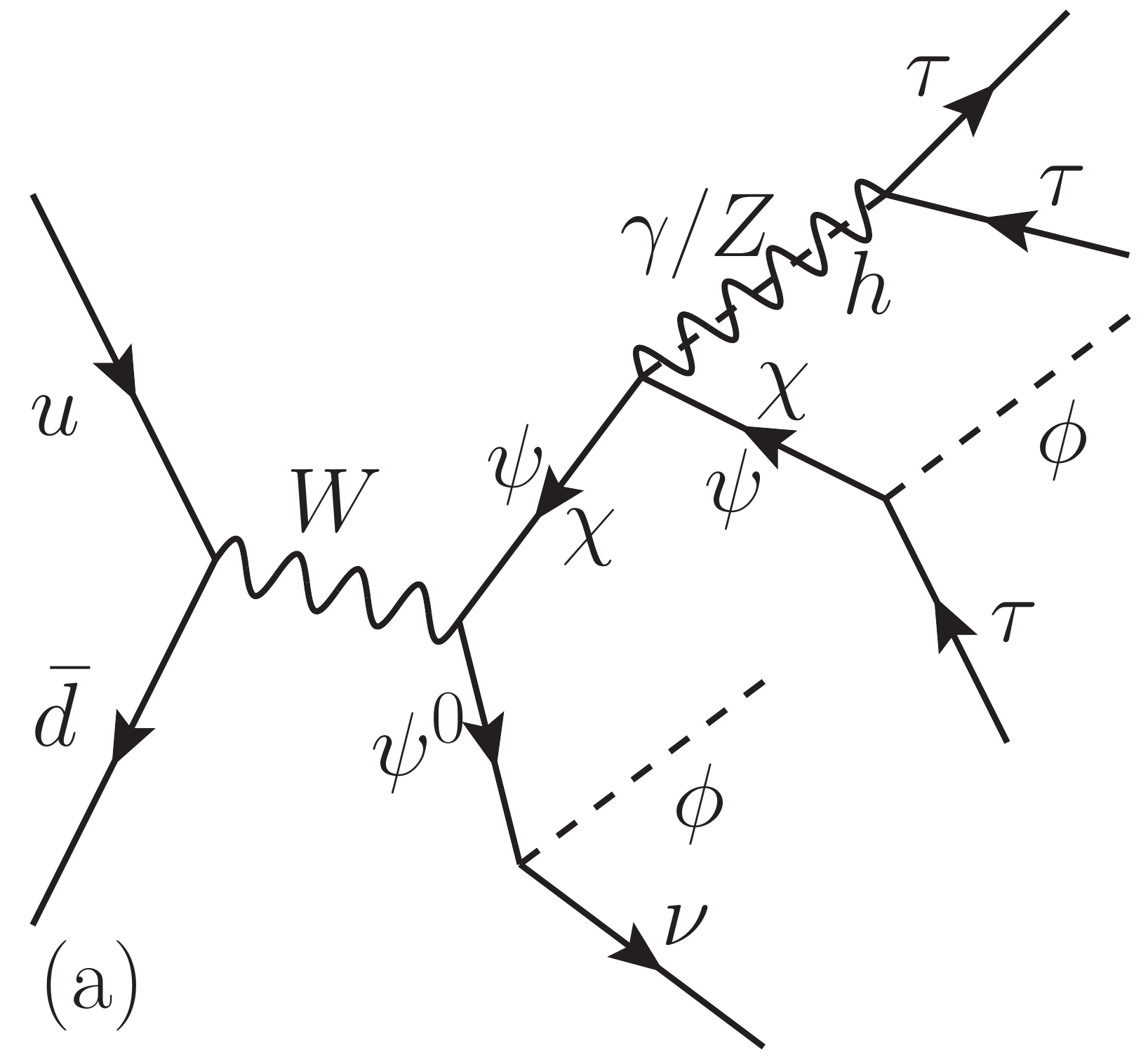}
 \hskip4em
 \includegraphics[height=0.2\textwidth]{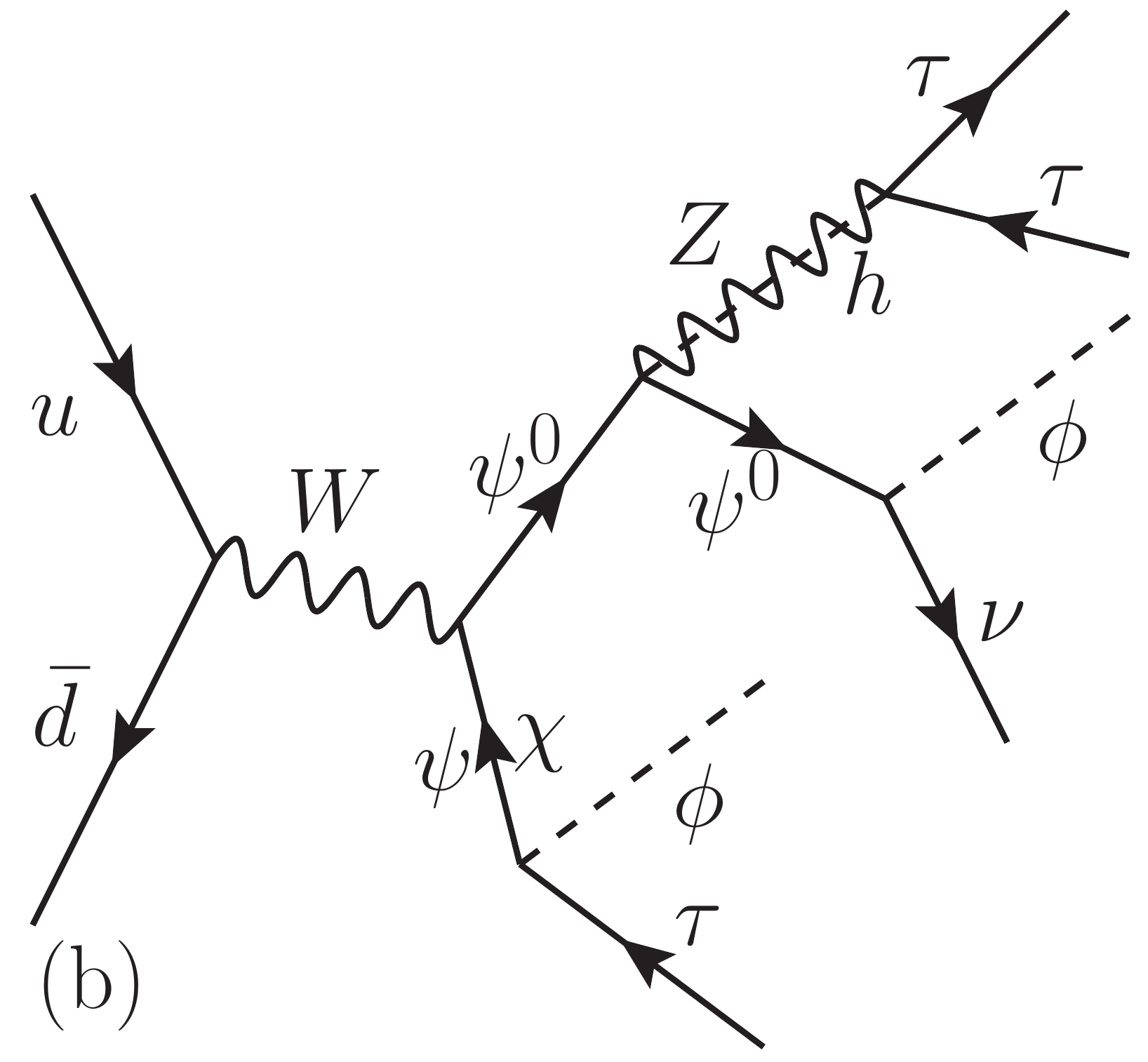}
 \hskip4em
 \includegraphics[height=0.2\textwidth]{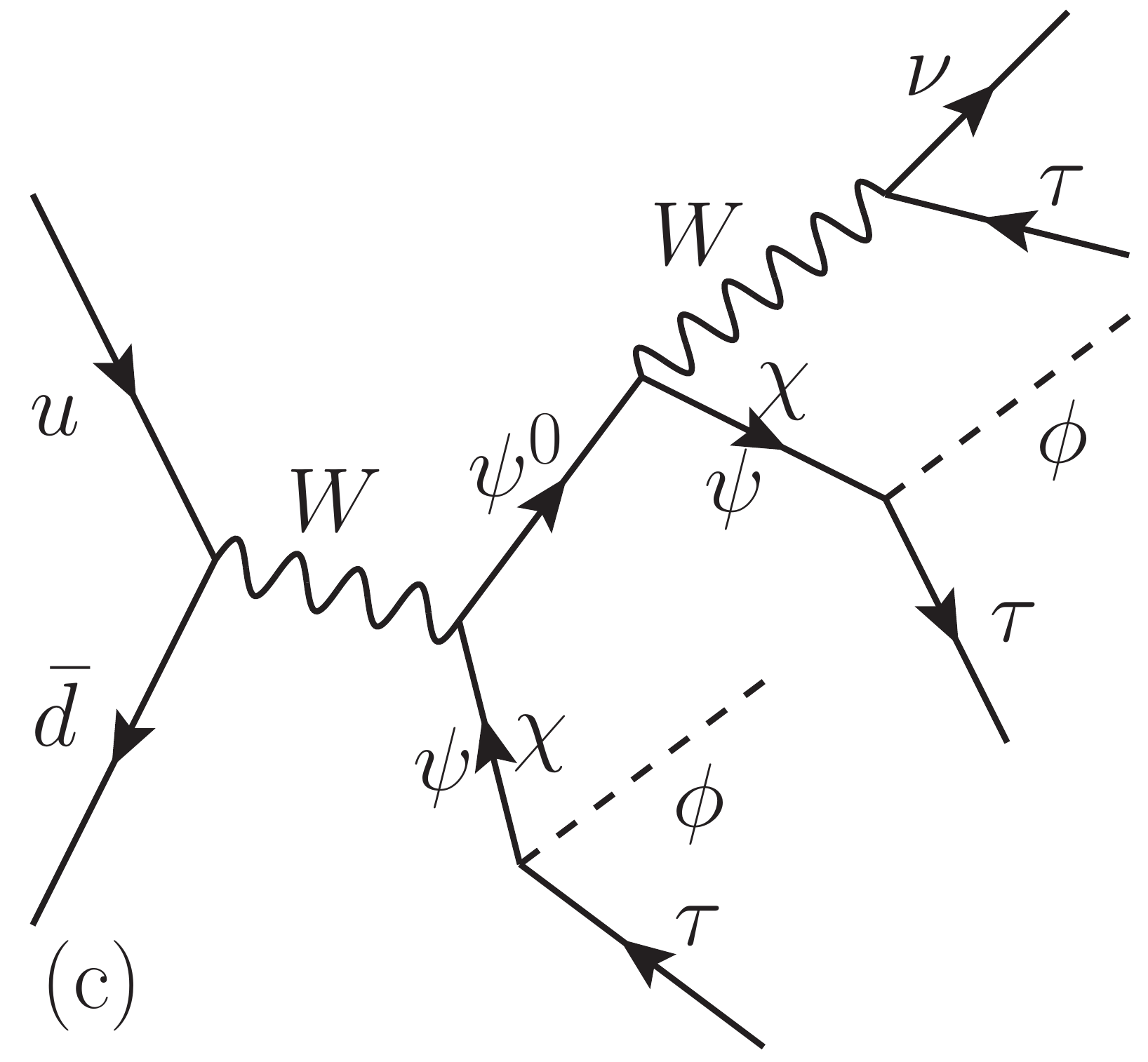}
 \\\vskip1em
 \includegraphics[height=0.2\textwidth]{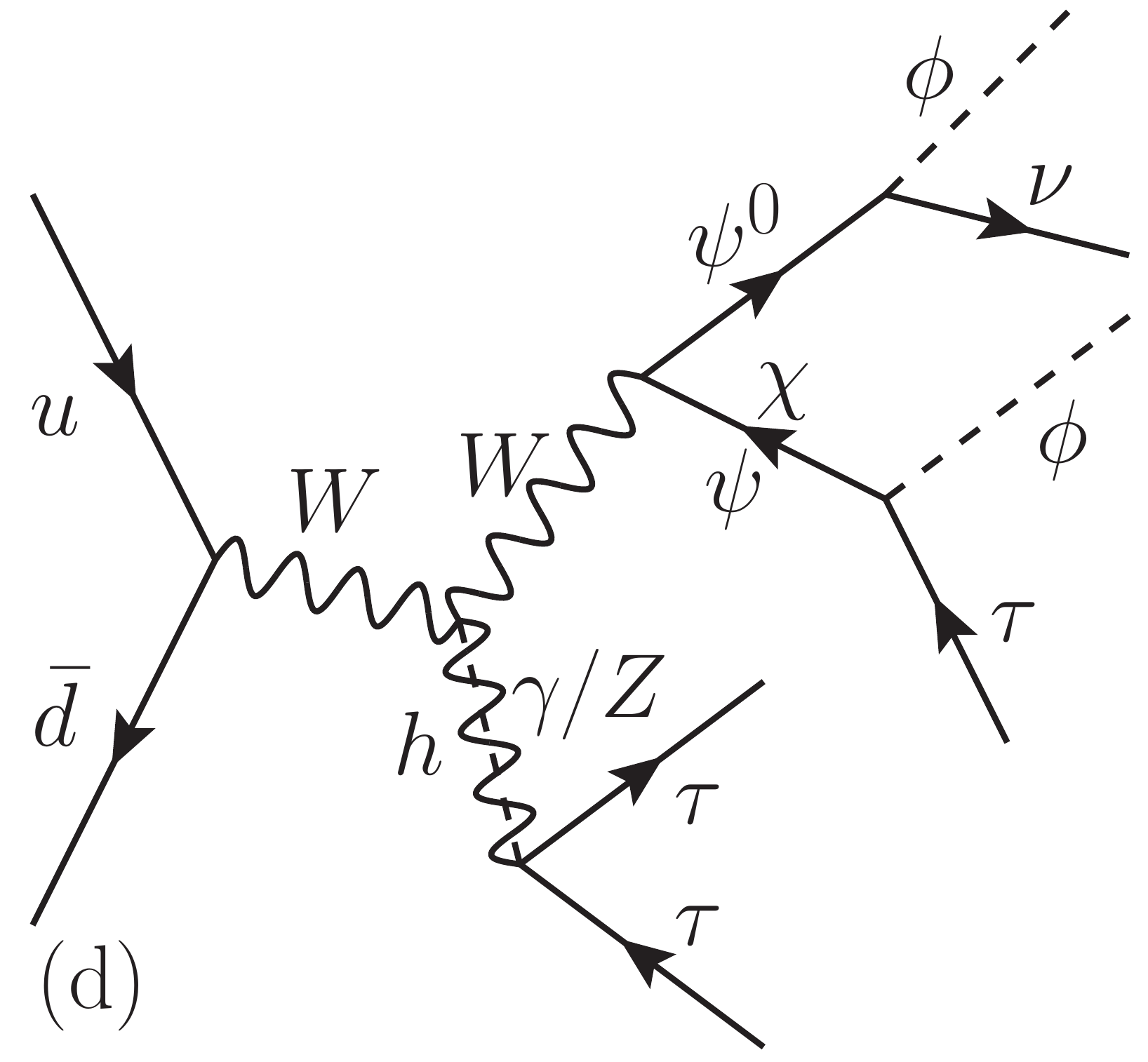}
 \hskip2em
 \includegraphics[height=0.2\textwidth]{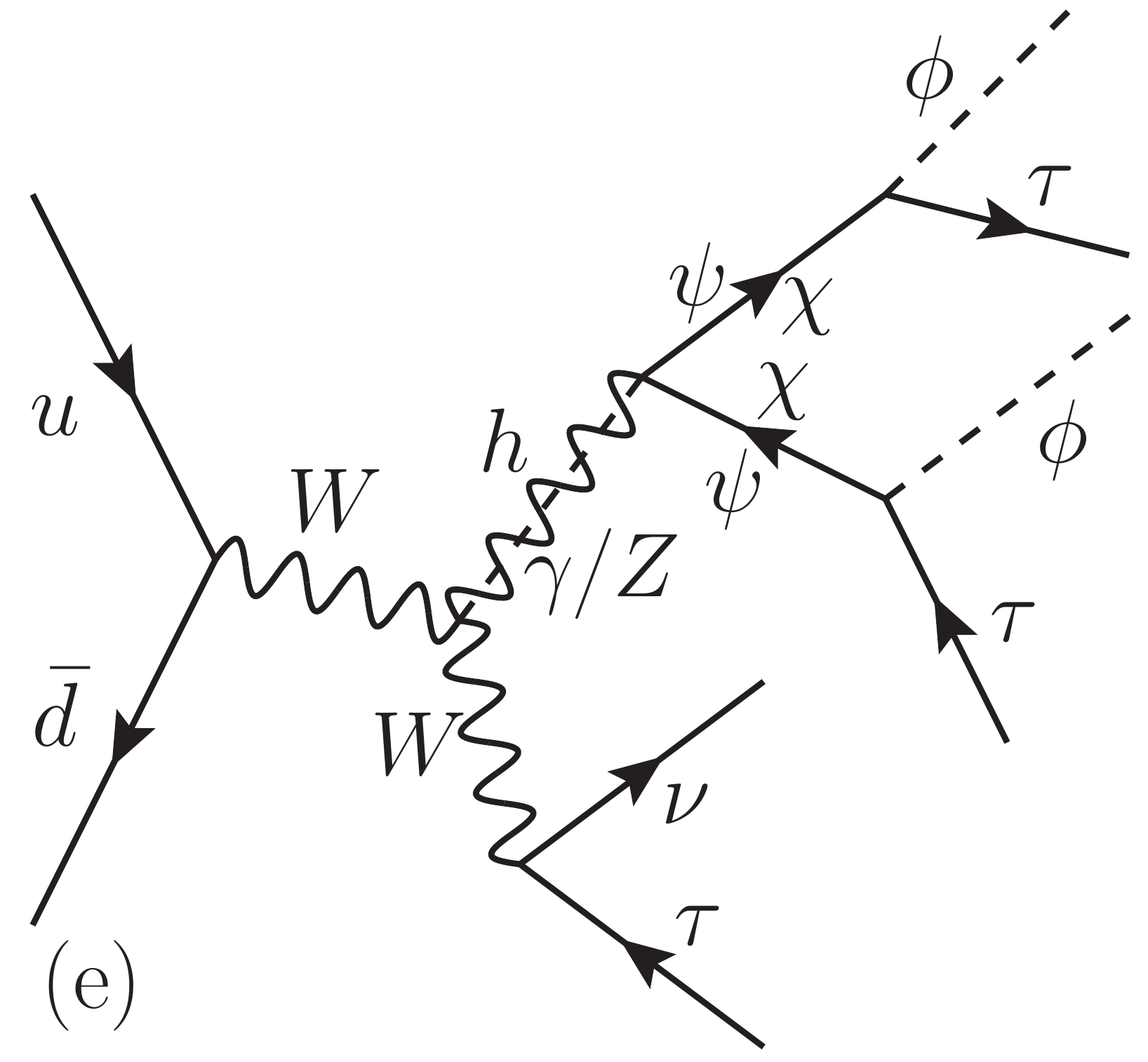}
 \hskip2em
 \includegraphics[height=0.2\textwidth]{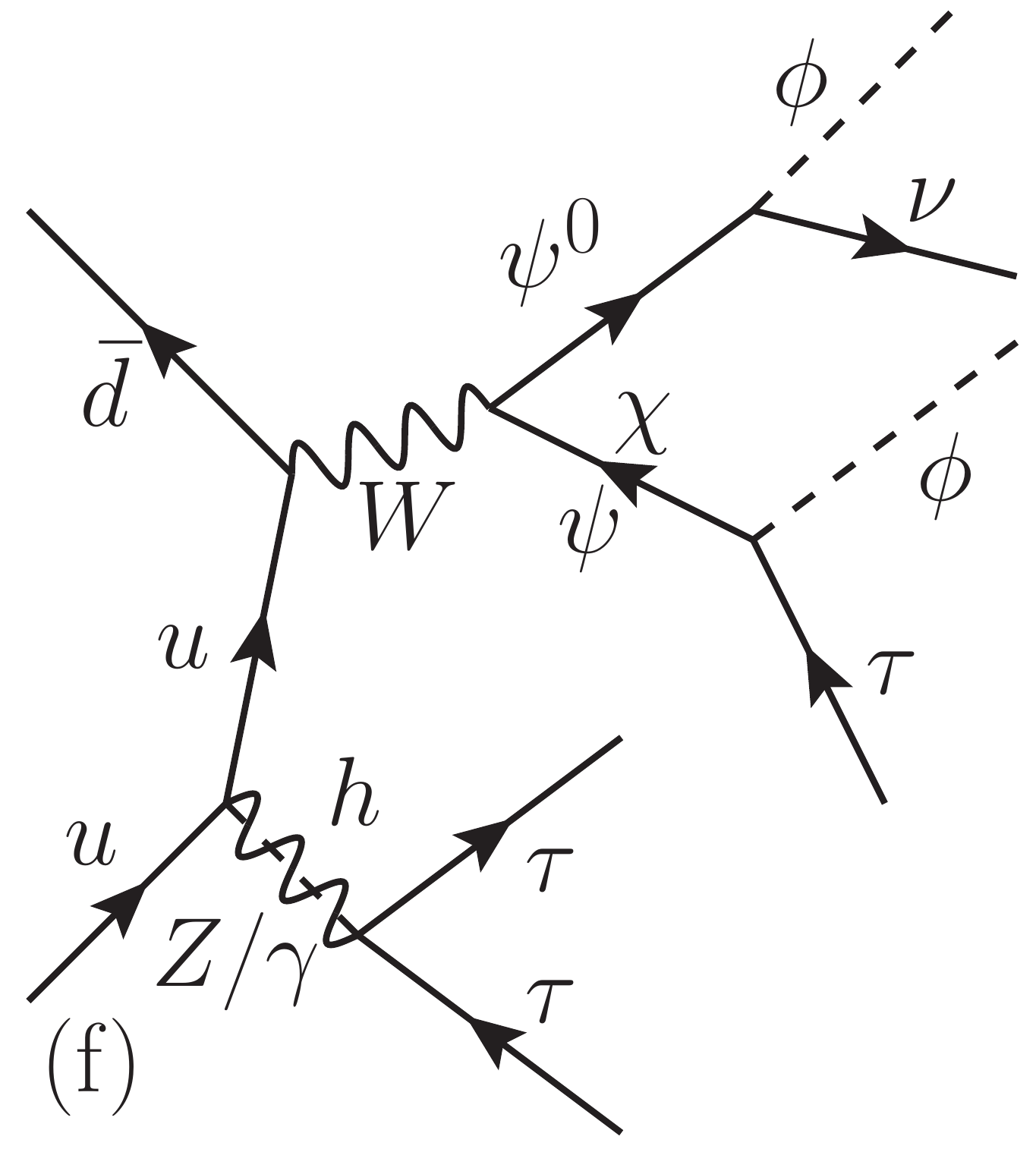}
 \hskip2em
 \includegraphics[height=0.2\textwidth]{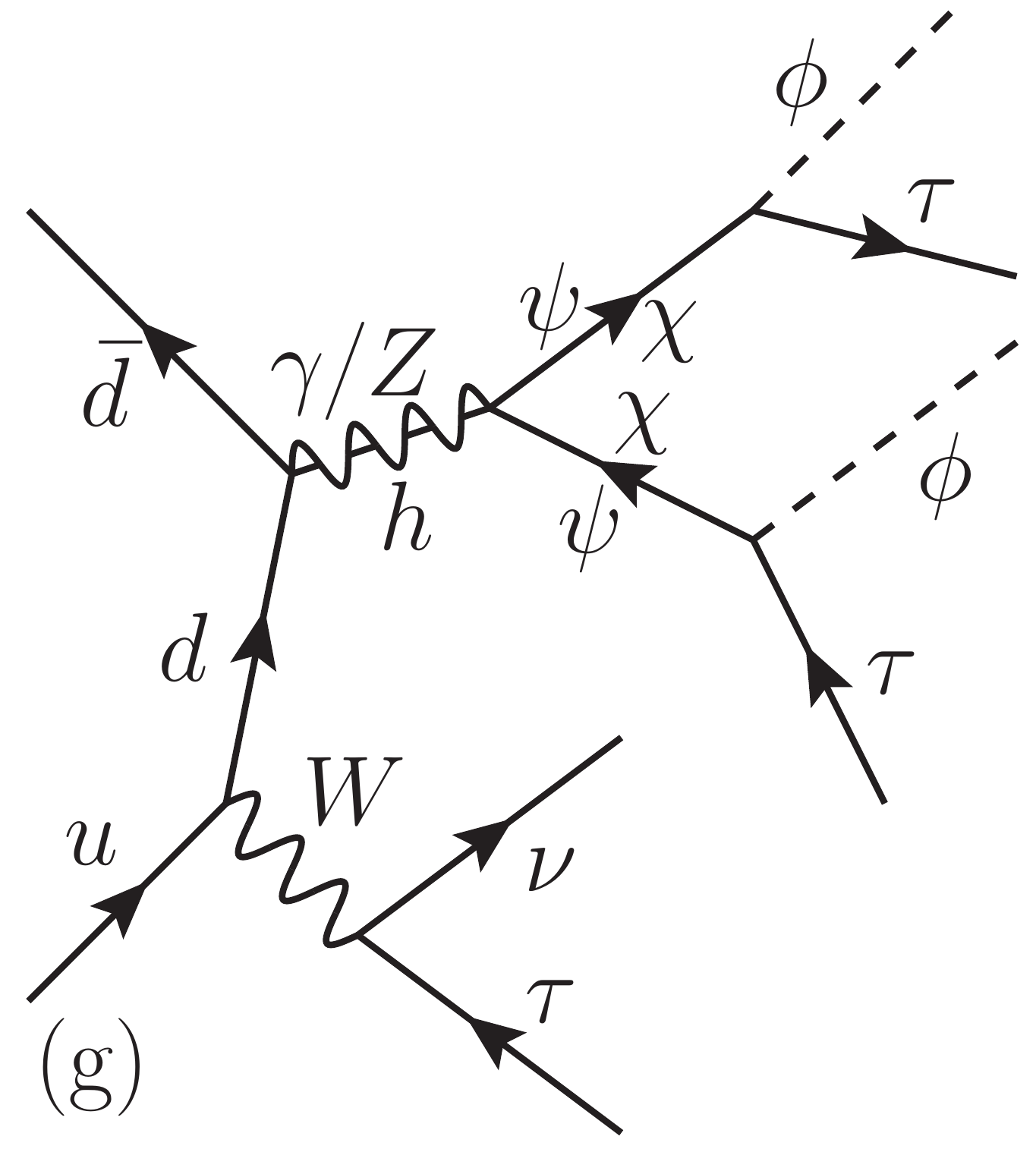}
 \caption{Feynman diagrams contributing to the \(3\tau\,\nu\, 2\phi\) channel at the LHC. (i) The upper panel shows the $W$-mediated $s$-channel processes through $q\widebar q^\prime \to \psi^0\,\psi^+ (\chi^+)$ channel; and (ii) the lower panel, $W$-mediated $s$-channel as well as quark-mediated $t$-channel processes through $q\widebar q^\prime \to W\,Z (\gamma)$ and $q\widebar q^\prime \to W\,h$ channels. The diagrams of the lower panel are seen to be suppressed, which makes the \(3\tau\,\nu\, 2\phi\) channel predominantly $W$-mediated.}
 \label{diag:3tau}
\end{figure}

From \cref{FR_Gauge}, we can see that the couplings of $W$ boson with the dark fermions are functions of mixing angle $\alpha$. Without the mixing, i.e., when $s_\alpha=0$, $\psi^+$ is exclusively a doublet and $\chi^+$ purely a singlet. This fact immediately implies that when $W^+ \to \psi^0\chi^+$ channels are off, the resulting final states are a consequence of pure doublet contribution. Hence, by tuning the mixing, one can control the \% of singlet contribution in the channels. The channels we can look into are as follows:

\begin{description}[leftmargin=0pt,labelindent=0pt]
 \item{\(\bm{\tau\,\nu\,2\phi}\)} : $s$-channel processes via $W^+ \to \psi^0\,\psi^+ (\chi^+)$ followed by $\psi^0$ decaying into $\nu\,\phi$, which remains invisible, and $\psi^+ (\chi^+) \to \phi\,\tau^+$. Since the only visible final state in this channel is a single $\tau$, and the missing energy can come from both $\psi^0$ and $\psi^+ (\chi^+)$, it is difficult to conclude anything about the DM signature.

 \item{\(\bm{3\tau\,\nu\, 2\phi}\)} : Here, the signal processes can proceed through the following different modes:
 \begin{enumerate}[(i)]
   \item $W$-mediated $s$-channel processes through $q\widebar q^\prime \to \psi^0\,\psi^+ (\chi^+)$ channel; and
   \item $W$-mediated $s$-channel as well as quark-mediated $t$-channel processes through $q\widebar q^\prime \to W\,Z (\gamma)$ and $q\widebar q^\prime \to W\,h$ channels.
 \end{enumerate}
 We encounter the latter case in the context of the {\em Unitarity problems} of the gauge bosons and is very similar to the $f\widebar f^\prime \to W\, Z (\gamma)$ and $f\widebar f^\prime \to W\,h$. See for example Refs. \cite{Choudhury:2012tk,Dahiya:2013uba,Ghosh:2017coz} for some recent papers in this context and the references therein. The point is that as the CM energy increases, the cross section of these channels decreases, which we can explain from the {\em Equivalence Theorem} of the gauge bosons. We see a similar trend in our case also. As a result, the dominant channels of \(3\tau\,\nu\, 2\phi\) process become mostly $W$-mediated $s$-channels mentioned in point (i) above. The diagrams mentioned in point (ii) are suppressed to point (i). See \cref{diag:3tau} for the complete diagrams with their decay channels.
\end{description}

\begin{figure}[!ht]
 \centering
 \includegraphics[width=0.4\textwidth]{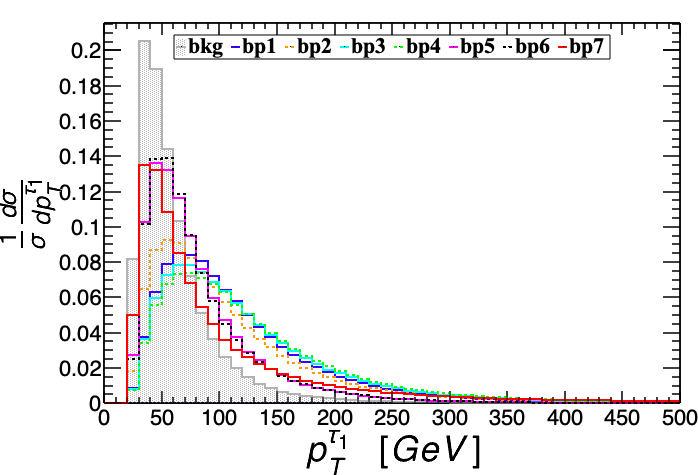}
 \includegraphics[width=0.4\textwidth]{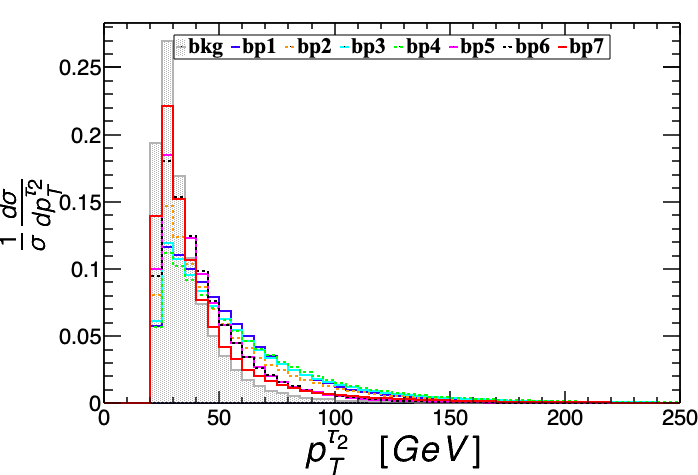}
 \\
 \includegraphics[width=0.4\textwidth]{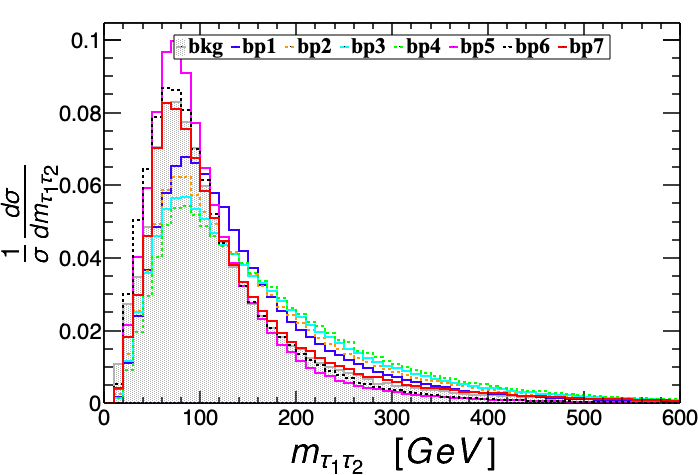}
 \includegraphics[width=0.4\textwidth]{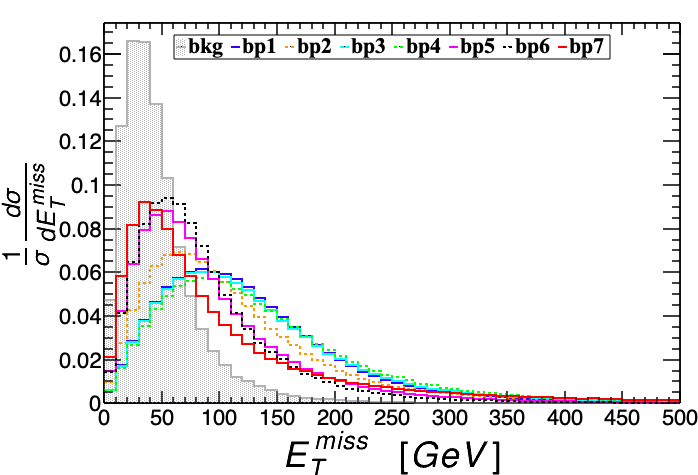}
 \\
 \includegraphics[width=0.4\textwidth]{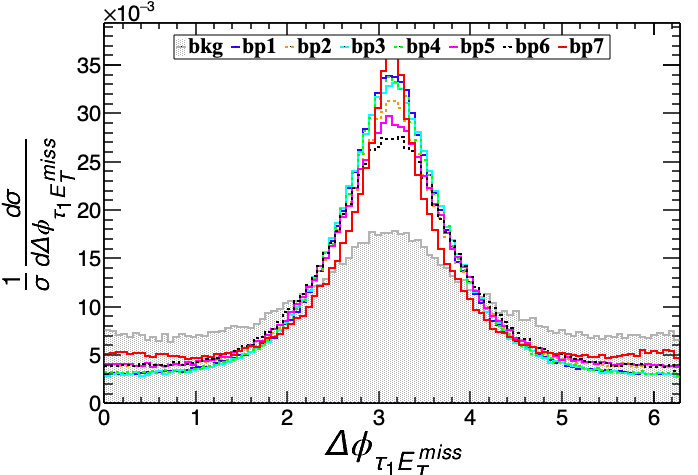}
 \includegraphics[width=0.4\textwidth]{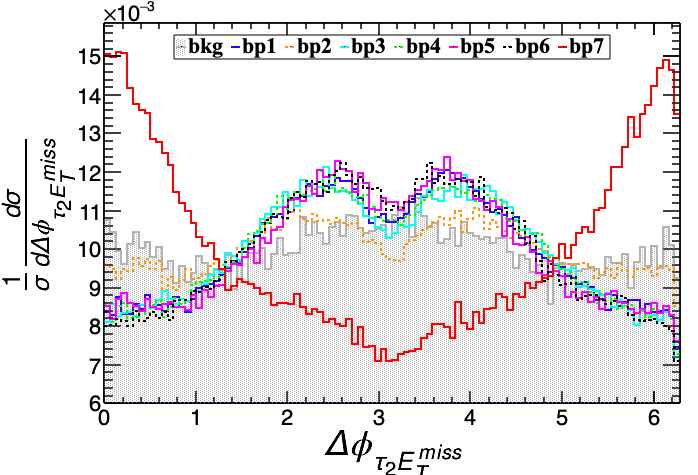}
 \caption{Significant distributions of the \(3\tau\,\nu\, 2\phi\) channel. Gradual shift of the peaks in the top and middle panel can be attributed to the change in the mixing parameter. $\Delta\phi$ distributions shows direction of the production of missing energy signal w.r.t. $\tau$'s, which are clearly distinguishable from the background.}
 \label{fig:3tau}
\end{figure}

In the CM frame, the two particles split with equal and opposite transverse momentum ($p_T$), whose magnitude is, say $p^{CM}_T$. As we can see from \cref{diag:3tau}, one of these two particles decays into two-body final states, whereas the other one splits into four-body final states, resulting into 3$\tau$'s and missing energy signal. If we distinguish these three $\tau$ leptons according to the decreasing order of their $p_T$, the leading-$p_T$ $\tau$ will mostly come from the two-body decays, i.e., \cref{diag:3tau}b) and (c). On the other hand, the two subleading $\tau$'s will probably come from the four-body decays mentioned above. As the third $\tau$ is the least energetic one, it will be difficult to explain the kinematics of the third $\tau$ with multiple accompanying invisible particles coming from such four-body decays. Moreover, the statistics of the third $\tau$ also comes out to be less than the other two $\tau$'s\footnote{This assertion is purely based on our Monte Carlo data sample. We showed, in \cref{fig:3tau}, some representative distributions which can be useful in depicting the effects of mixing. To make a conclusive statement about the features of the third $\tau$ in this process, we need a very large data set of Monte Carlo events. This difficulty is more so because of $\tau$ detection efficiency at the LHC.}. Hence we will confine our discussions within the two leading-$p_T$ $\tau$'s and their observables. It is obvious from \cref{diag:3tau} that, the two-body visible channel is $\psi^+ (\chi^+) \to \phi\,\tau^+$ (\cref{diag:3tau}b) and (c)), whereas the invisible channel is $\psi^0 \to \nu\,\phi$ (\cref{diag:3tau}a)). In \cref{fig:3tau}, we show some of the relevant distributions of the $3\tau$ channel for a set of benchmark points (see \cref{tab:bp_3tau}) which satisfy the required relic density. Here the relic density is satisfied through pair annihilation channels as we concluded for the doublet case \cite{Chakraborti:2019fnz}.

\begin{table}[!ht]
 \centering
 {\setlength{\tabcolsep}{1em}
 \begin{tabular}
 { c | r >{\columncolor{Gray}}r r | c | l | l >{\columncolor{Gray}}l l | c }
 \hline
     &$m_\phi$
             & \multicolumn{1}{ c }{\cellcolor{Gray}$m_{\psi^0}$}
                     & \multicolumn{1}{ c }{$m_{\psi^\pm}$}
                             & $m_\chi$
                                    & \multicolumn{1}{ c }{\multirow{2}*{$y_\tau^D$}}
                                           & \multicolumn{1}{ c }{\multirow{2}*{$y_\tau^S$}}
                                                  &       & \multicolumn{1}{ c |}{\multirow{2}*{$s_\alpha$}}
                                                                 & cross section \\
     & \multicolumn{4}{ c|}{[ GeV ]}& \multicolumn{1}{ c }{}
                                           &      & \multicolumn{1}{ c }{\cellcolor{Gray}\multirow{-2}*{$y$}}
                                                          &      & [ fb ]    \\
 \hline
 \hline
 BP1 & $100$ & $348$ & $350$ &$180$ &$0.4$ &$2.05$&$0.097$&$0.1$ & $1.2$     \\
 \hline
 BP2 & $120$ & $249$ & $250$ &$150$ &$2.15$&$0.0$ &$0.057$&$0.1$ & $0.15$    \\
 \hline
 BP3 & $180$ & $358$ & $380$ &$295$ &$0.34$&$2.0$ &$0.2$  &$0.5$ & $0.10$    \\
 \hline
 BP4 & $180$ & $358$ & $380$ &$295$ &$0.72$&$0.85$&$0.2$  &$0.5$ & $0.058$   \\
 \hline
 BP5 &  $80$ & $250$ & $350$ &$150$ &$1.87$&$0.0$ &$0.57$ &$0.71$& $2.1$     \\
 \hline
 BP6 &  $80$ & $250$ & $350$ &$150$ &$0.55$&$2.5$ &$0.57$ &$0.71$& $12.2$    \\
 \hline
 BP7 &  $88$ &$1406$ &$1500$ &$110$ &$2.8$ &$0.0$ &$2.0$  &$0.26$& $0.049$   \\
 \hline
 \end{tabular}
 }
 \caption{Relic density allowed benchmark points for analysis of \(3\tau\,\nu\, 2\phi\) channels. $y_\ell^D=y_\ell^S=10^{-9}$. The shaded columns are for the dependent model parameters.}
 \label{tab:bp_3tau}
\end{table}

From the distributions of \cref{fig:3tau}, we can deduce the following inferences:

 (i) We see sharp ``Jacobian peaks'' in the $p_T$ distributions for leading and subleading-$p_T$ $\tau$'s for BPs 5-7. These ``Jacobian peaks'' appear at $m/2$ in $p_T$-distributions and at $m$ in missing transverse mass distributions. Here $m$ is the mass of the parent particle which decays to $\tau$. From the peaks of \cref{fig:3tau}a) and (b) we can infer that $\tau_1$ is coming from $\chi$ and $\tau_2$ from $W$ bosons for BP7. We can conclude that both $\chi$ and $W$ boson are on-shell for this benchmark point. For BPs 5-6, the peaks are slightly towards the right. This fact signifies that here the leading and sub-leading $\tau$'s do not entirely come from $\chi$ and $W$, but predominantly so. We see further confirmation for this from the distributions of the respective transverse masses which we have not included for the sake economy of space.

 (ii) The BPs 1-4 and 5-7 can be similarly classified. While the earlier set gives a relatively flatter profile, the latter shows sharp peaks in the $p_T$ and $E^{miss}_T$ distribution plots. From the discussion above, we can see that the reason for this is the off-shell-ness of the mediating particles. This point is also clear from the values of the parameters in \cref{tab:bp_3tau}.
  
 (iii) The $p^{\tau_1}_T$ distribution profile is similar to that of $E^{miss}_T$, whereas the peak of $\Delta\phi_{\tau_1 E^{miss}_T}$ shows that $\tau_1$ and $E^{miss}_T$ are back to back in nature. We can infer that they are coming from the same parent particles and so they are equal and opposite in nature.
  
 (iv) The invariant mass plot of $\tau_1$ and $\tau_2$ gives the signal that they are pair produced from the $Z$ boson decay.
  
 (v) The distinct features of BP 7 are evident from all the distributions. The very high values of $m_{\psi^{0\pm}}$ indicate that these particles in the intermediate states will suppress the effects of the respective diagram very much compared to $\chi$.

\subsubsection{\texorpdfstring{\(\ell\,\tau + E^{miss}_T\)}{ell tau + MET} channel}
Now we shall look into the effects of mixing in the $Z$ boson mediated signal processes. For this, we focus on the \(\ell\,\tau\,2\nu\, 2\phi\) channel. The processes in this channel can proceed through the following different modes, as can be seen from \cref{diag:1tau1lep}:
\begin{figure}[!ht]
 \centering
 \includegraphics[height=0.2\textwidth]{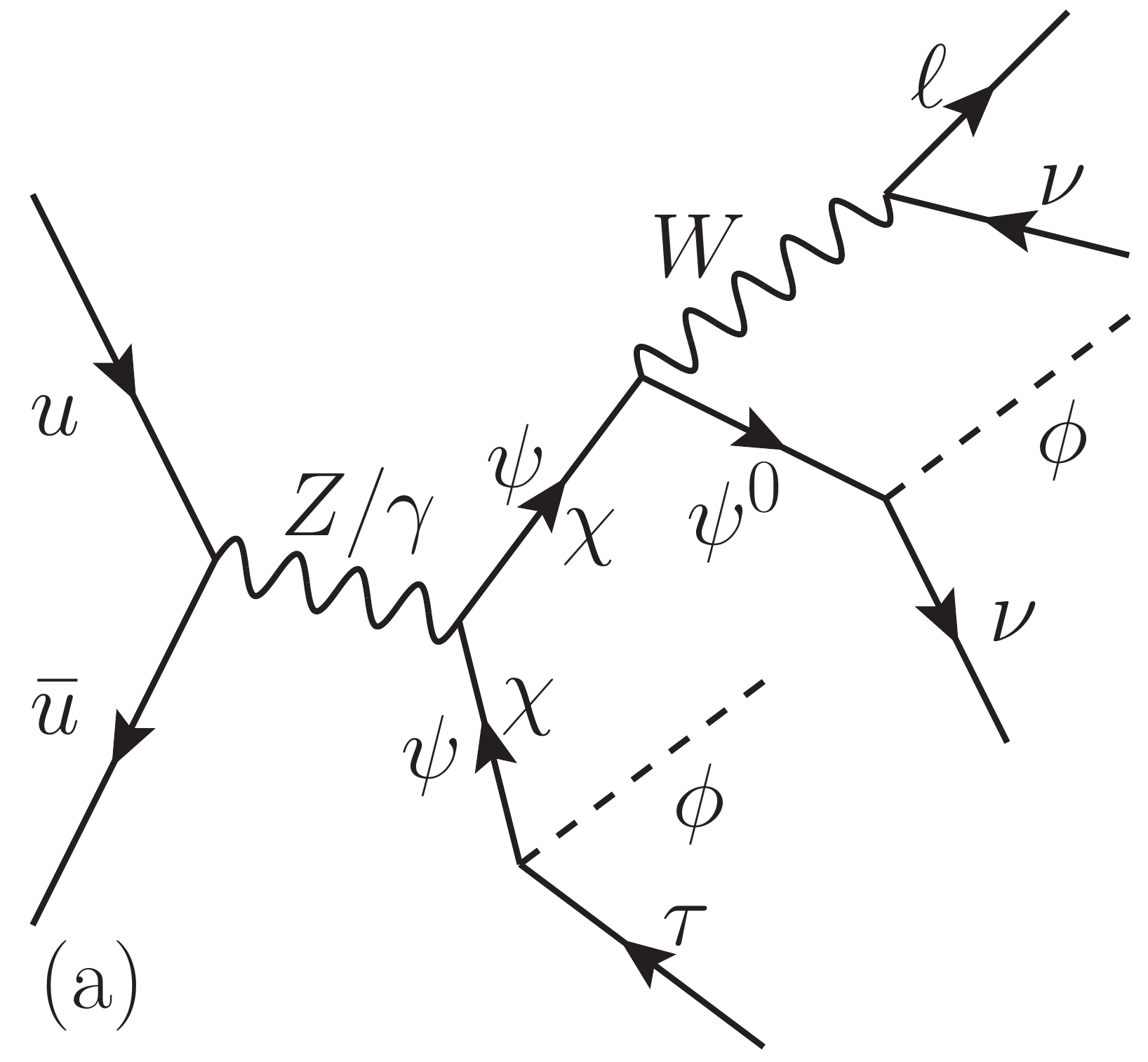}
 \hskip4em
 \includegraphics[height=0.2\textwidth]{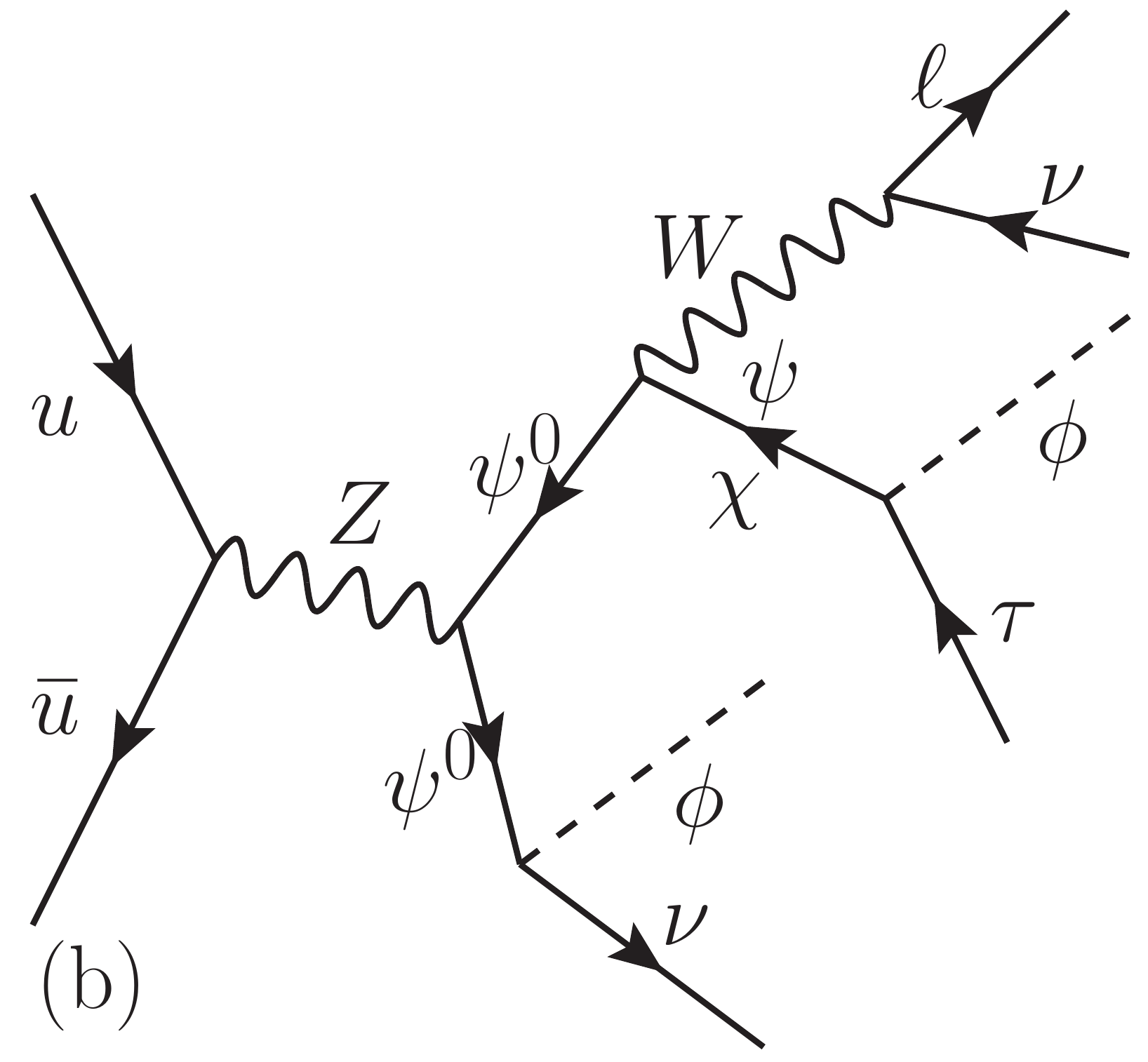}
 \\\vskip1em
 \includegraphics[height=0.2\textwidth]{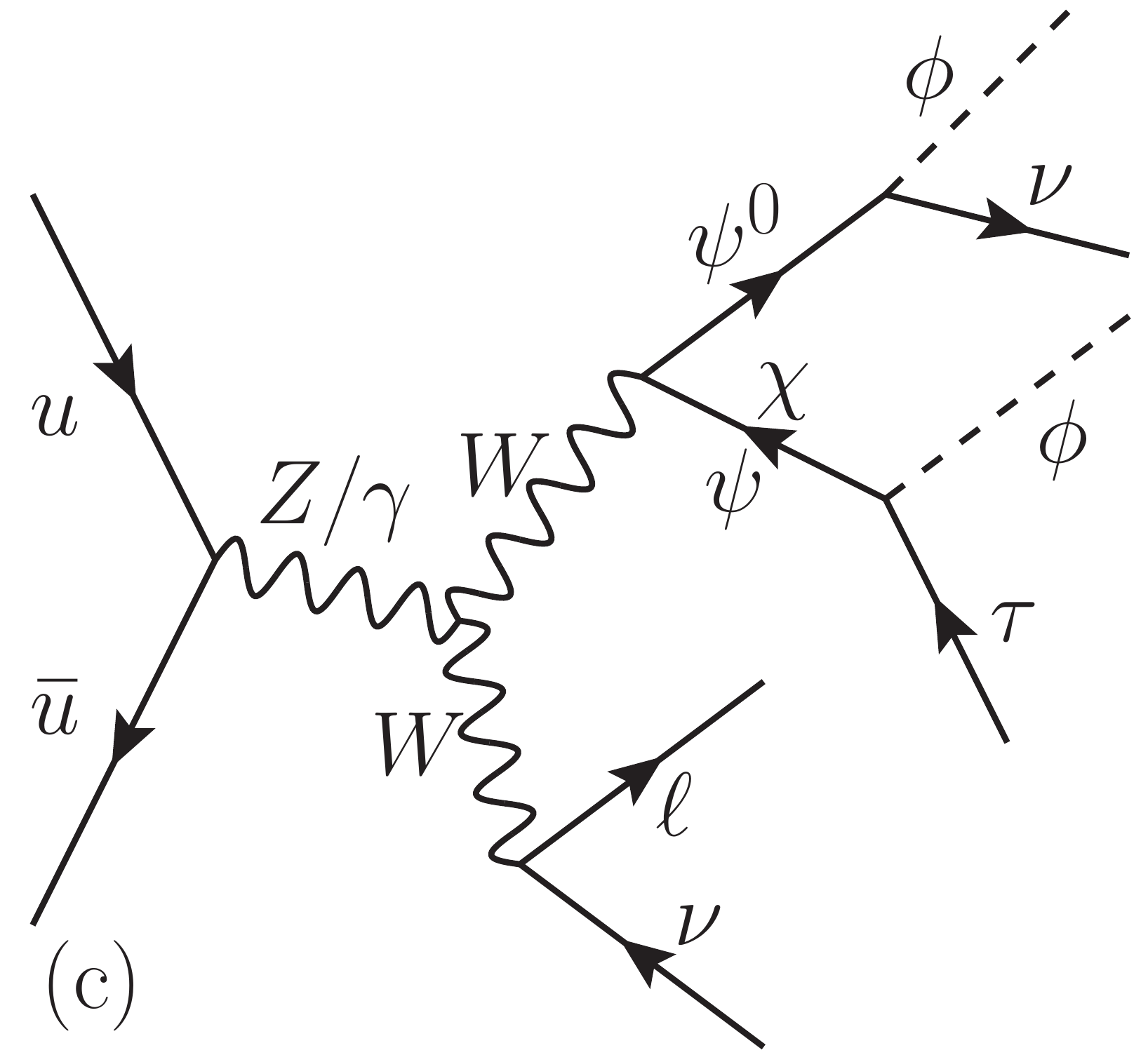}
 \hskip7em
 \includegraphics[height=0.2\textwidth]{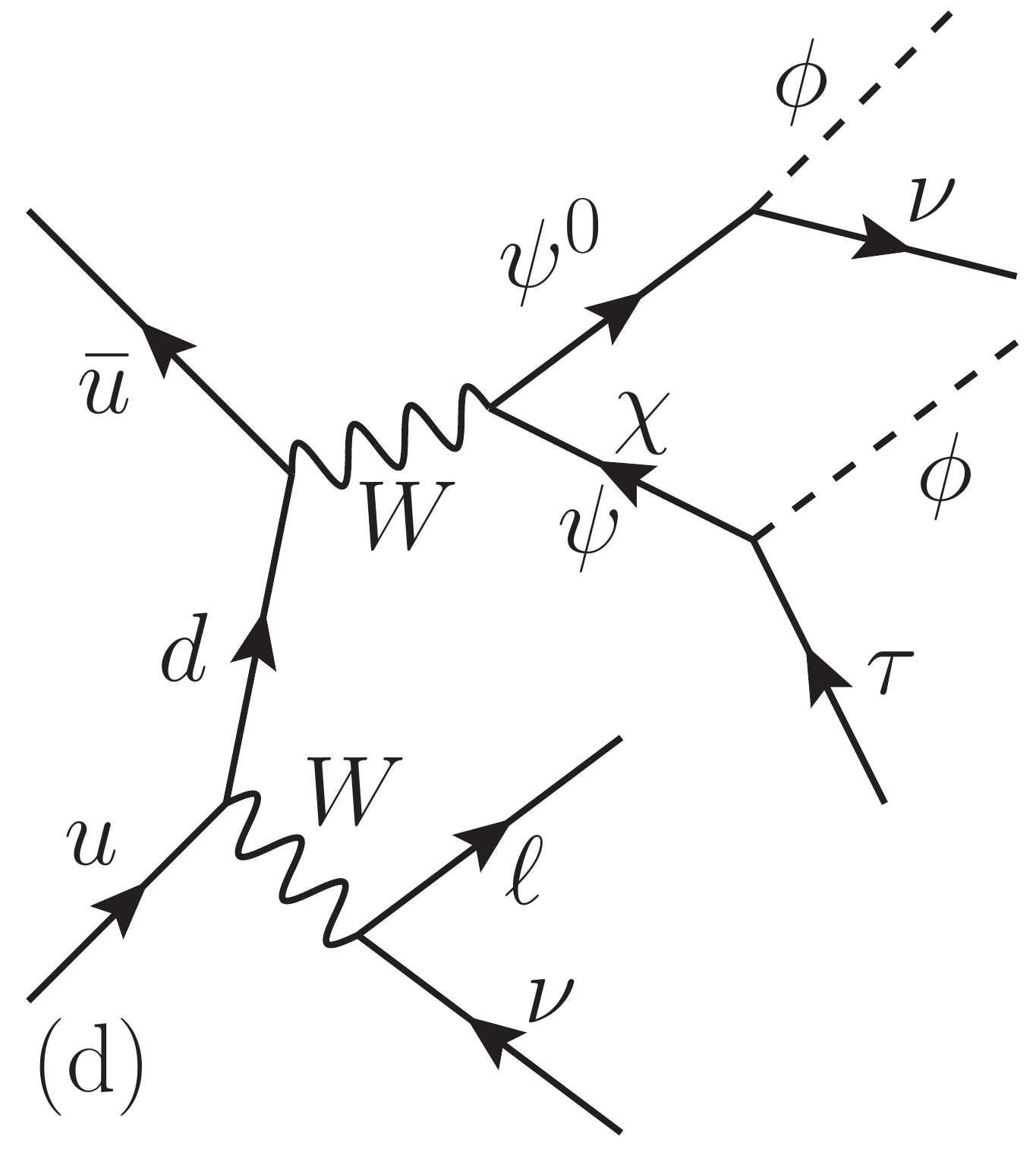}
 \caption{Feynman diagrams contributing to the \(\ell\,\tau\,2\nu\, 2\phi\) channel at the LHC. (i) The upper panel shows the $Z$-mediated $s$-channel processes through $q\widebar q \to \psi^+\,\psi^-, \chi^+\,\chi^-, \psi^+\,\chi^-$ and $\psi^0\widebar\psi^0$ channel; and (ii) the lower panel, $Z$-mediated $s$-channel as well as quark-mediated $t$-channel processes through $q\widebar q \to W\,W$ channel. The diagrams of the lower panel are seen to be suppressed, which makes the \(\ell\,\tau\,\nu\, 2\phi\) channel predominantly $Z$-mediated.}
 \label{diag:1tau1lep}
\end{figure}
\begin{enumerate}[(i)]
  \item $Z$ boson mediated $s$-channel processes through $q\widebar q \to \psi^+\,\psi^-, \chi^+\,\chi^-, \psi^+\,\chi^-$ and $\psi^0\widebar\psi^0$ channel; and
  \item $Z$ boson mediated $s$-channel as well as quark-mediated $t$-channel processes through $q\widebar q \to W\,W$ channel.
\end{enumerate}
Among these, the processes in point (i) are much dominant to the processes in point (ii). This is because, as in the case for \(3\tau\,\nu\, 2\phi\) channel, here also the latter case is very similar to the $f\widebar f \to W\,W$ in the context of the Unitarity problems of the gauge bosons \cite{Choudhury:2012tk,Dahiya:2013uba,Ghosh:2017coz}. Hence, as the CM energy increases, the cross section of these channels decreases, that can be explained from the Equivalence Theorem of the gauge bosons. Among the remaining diagrams of point (i), the $\gamma$ mediated processes are independent of the mixing parameter $s_\alpha$ at the production level, as can be seen from \cref{FR_Gauge}. As a result, they are not interesting for our purpose, where we focus on the effects of the mixing parameter in the distributions. The \(\ell\,\tau\,2\nu\, 2\phi\) channel can also have contributions from $gg$ initiated $s$-channel processes which are very much suppressed for the chosen benchmark points (see \cref{tab:bp_1tau1lep}). The $q\widebar q$ initiated Higgs boson channels are also suppressed due to negligible $hq\widebar q$ couplings. So, it should be sufficient to focus only on the processes in point (i) above, while explaining the features of this channel. In \cref{fig:1tau1lep}, we show the relevant distributions of the \(\ell\,\tau\,2\nu\, 2\phi\) channel for the set of benchmark points in \cref{tab:bp_1tau1lep} which satisfy the required relic density.

\begin{table}[!ht]
 \centering
 {\setlength{\tabcolsep}{1em}
 \begin{tabular}
 { r@{} @{}l | c >{\columncolor{Gray}}l c | c | c | l >{\columncolor{Gray}}l l | c }
 \hline
   &     & $m_\phi$
              & \multicolumn{1}{ c }{\cellcolor{Gray}$m_{\psi^0}$}
                      & \multicolumn{1}{ c }{$m_{\psi^\pm}$}
                           & $m_\chi$
                                & \multicolumn{1}{ c }{\multirow{2}*{$y_\tau^D$}}
                                     & \multicolumn{1}{ c }{\multirow{2}*{$y_\tau^S$}}
                                             &         & \multicolumn{1}{ c |}{\multirow{2}*{$s_\alpha$}}
                                                              & cross section \\
   &     & \multicolumn{4}{ c|}{[ GeV ]}
                                & \multicolumn{1}{ c }{}
                                     &       & \multicolumn{1}{ c }{\cellcolor{Gray}\multirow{-2}*{$y$}}
                                                       &      & [ fb ]        \\
 \hline
 \hline
 BP &    & \multirow{6}*{$120$}
              & $300$ & \multirow{6}*{$400$}
                           & \multirow{6}*{$200$}
                                & \multirow{6}*{$0.5$}
                                     & $0.23$& $0.57$  & $1/\surd{2}$
                                                              & $4.4$         \\
 \hhline{>{\arrayrulecolor{black}}==~=~|~|~:====}
 BP &\_5 &    & $350$ &    &    &    & $0.26$& $0.5$   &$0.5$ & $1.88$        \\
 \hhline{>{\arrayrulecolor{black}}--~-~|~|~|----}
 BP &\_1 &    & $398$ &    &    &    & $1.55$& $0.11$  &$0.1$ & $0.73$        \\
 \hhline{>{\arrayrulecolor{black}}--~-~|~|~|----}
 BP &\_05&    &$399.5$&    &    &    & $2.2$ & $0.0574$&$0.05$& $0.35$        \\
 \hhline{>{\arrayrulecolor{black}}--~-~|~|~|----}
 BP &\_03&    &$399.82$
                      &    &    &    & $2.12$& $0.0345$&$0.03$& $0.25$        \\
 \hhline{>{\arrayrulecolor{black}}--~-~|~|~|----}
 BP &\_0 &    & $400$ &    &    &    & $1.82$& $0.0$   &$0.0$ & $0.18$        \\
 \hline
 \end{tabular}
 }
 \caption{Relic density allowed benchmark points for analysis of \(\ell\,\tau\,2\nu\, 2\phi\) channels. $y_\ell^D=y_\ell^S=10^{-9}$. The shaded columns are for the dependent model parameters.}
 \label{tab:bp_1tau1lep}
\end{table}

\begin{figure}[!ht]
 \centering
 \includegraphics[width=0.4\textwidth]{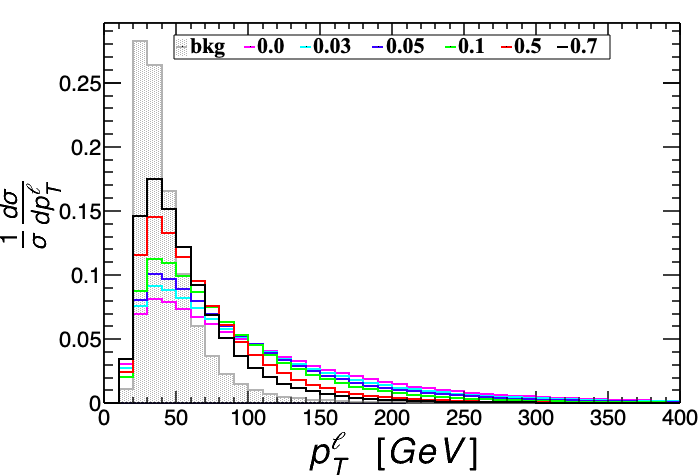}
 \includegraphics[width=0.4\textwidth]{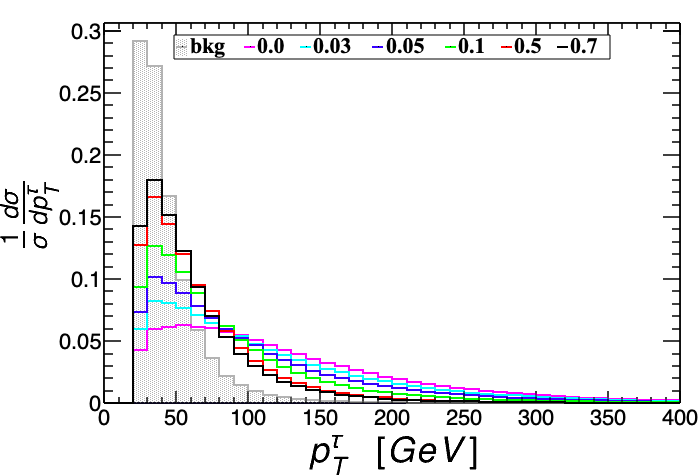}
 \\
 \includegraphics[width=0.4\textwidth]{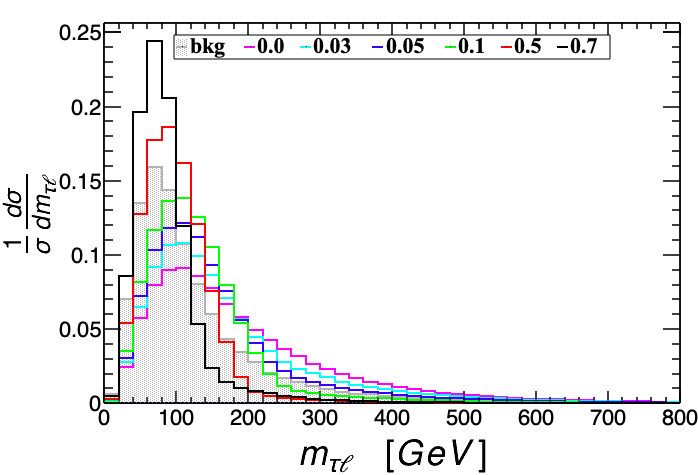}
 \includegraphics[width=0.4\textwidth]{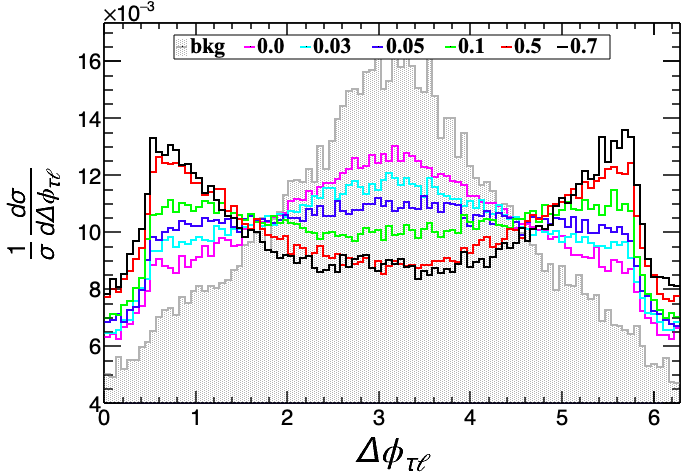}
 \caption{Some significant distributions of the \(\ell\,\tau\,2\nu\, 2\phi\) channel. As for the \(3\tau\,\nu\, 2\phi\) case, gradual shift of the peaks in the distributions can be attributed to the change in the mixing parameter. The distributions are labelled with the respective values of $s_\alpha$.}
 \label{fig:1tau1lep}
\end{figure}

Some interesting observations from the distributions of \cref{fig:1tau1lep}, are as follows:

 (i) If we closely look into the benchmark points for this channel in \cref{tab:bp_1tau1lep}, they are chosen such that the first one is for $s_\alpha = 1/\sqrt{2}$ where there is equal mixing between the doublet and singlet component of the dark fermions. We gradually change the value of $s_\alpha$ to zero. The rest of the dependent and independent parameters are chosen just to keep the relic density within the allowed limit.

 (ii) For $s_\alpha = 0$, there is no mixing, and hence $\psi^+$ is a pure doublet state whereas $\chi^+$ is purely singlet. Now since we see from \cref{FR_Gauge} that $W^+\chi^-\psi^0$ and $Z\psi^+\chi^-$ vertices do not exist for such a case, we have only the doublet contribution for BP\_0. As a result, the distributions are independent of the values of $m_\chi$ and $y_\tau^S$.

 (iii) From the numbers given in \cref{tab:bp_1tau1lep}, we see that the role of mixing dictates the dominant channel, and hence, the trends of invariant mass distribution. Thus, we conclude from these numbers that, \cref{diag:1tau1lep}b) dominates for the values of $s_\alpha$ around $1/\sqrt{2}$ leading to a $Z$-peak for $s_\alpha = 1/\sqrt{2}$, whereas, for other values of $s_\alpha$ \cref{diag:1tau1lep}a) dominates. This can also be seen in the distribution of $\Delta\phi_{\tau\ell}$ in \cref{fig:1tau1lep}, where we see that both $\tau$ and $\ell$ are more probable to be along the same direction around $s_\alpha = 1/\sqrt{2}$. $\Delta\phi_{\tau E^{miss}_T}$ further establishes this conclusion where we will find $\tau$ and $E^{miss}_T$ going in the opposite directions. We have not kept $\Delta\phi_{\tau E^{miss}_T}$ plot here to avoid redundancy.

 (iv) The $p_T$ distributions of \cref{fig:1tau1lep} justify the dominance of the channels of \cref{diag:1tau1lep} as a function of the mixing angle. In both $p^\ell_T$ and $p^\tau_T$ distributions, it is seen that towards the low $p_T$ region, \cref{diag:1tau1lep}b) dominated channels are more probable whereas the high $p_T$ region favours \cref{diag:1tau1lep}a) dominated channels. This could be attributed to the feature that in \cref{diag:1tau1lep}b) dominated channels, the energy share of both the light lepton and $\tau$ is less than the energy share in \cref{diag:1tau1lep}a). In \cref{diag:1tau1lep}b), both the lepton and $\tau$ are produced from the decay of a single $\psi^0$ whereas in \cref{diag:1tau1lep}a), they are produced from different mediators. This feature is more prominent in $p^\tau_T$  rather than $p^\ell_T$. This is because in \cref{diag:1tau1lep}a), the branching of the decay chain producing $\tau$ is comparatively less than what it is in \cref{diag:1tau1lep}b), making the difference in energy share of $\tau$ between \cref{diag:1tau1lep}a) and (b) dominated channels more prominent. The lepton, on the other hand, is produced through more branching in both the figures, which makes the effect of reduced energy share less prominent as \cref{diag:1tau1lep}b) takes over \cref{diag:1tau1lep}a) with the increase of mixing angle in the chosen benchmark points.

We observe the same trend through all the distributions of \cref{fig:3tau,fig:1tau1lep}. There is a gradual change in the peak and tail positions of the kinematic distributions with the variations of $s_\alpha$. This change is more prominent for the latter case as we have chosen the benchmark points as such. Here, apart from the mixing parameter, $s_\alpha$, all the other dependent and independent parameters of the model are tweaked to some extent so that they satisfy the required relic density. Despite these small tweaks, we can say from our observations that this feature of the distributions is solely dependent on $s_\alpha$ and not anything else. We confirmed our assertion by keeping the rest of the independent parameters same and varying only the value of $s_\alpha$.

\subsection{Indirect Search Prospects}
\label{sec:det}
\begin{figure}[!ht]
  \centering
  \includegraphics[height=0.2\textwidth]{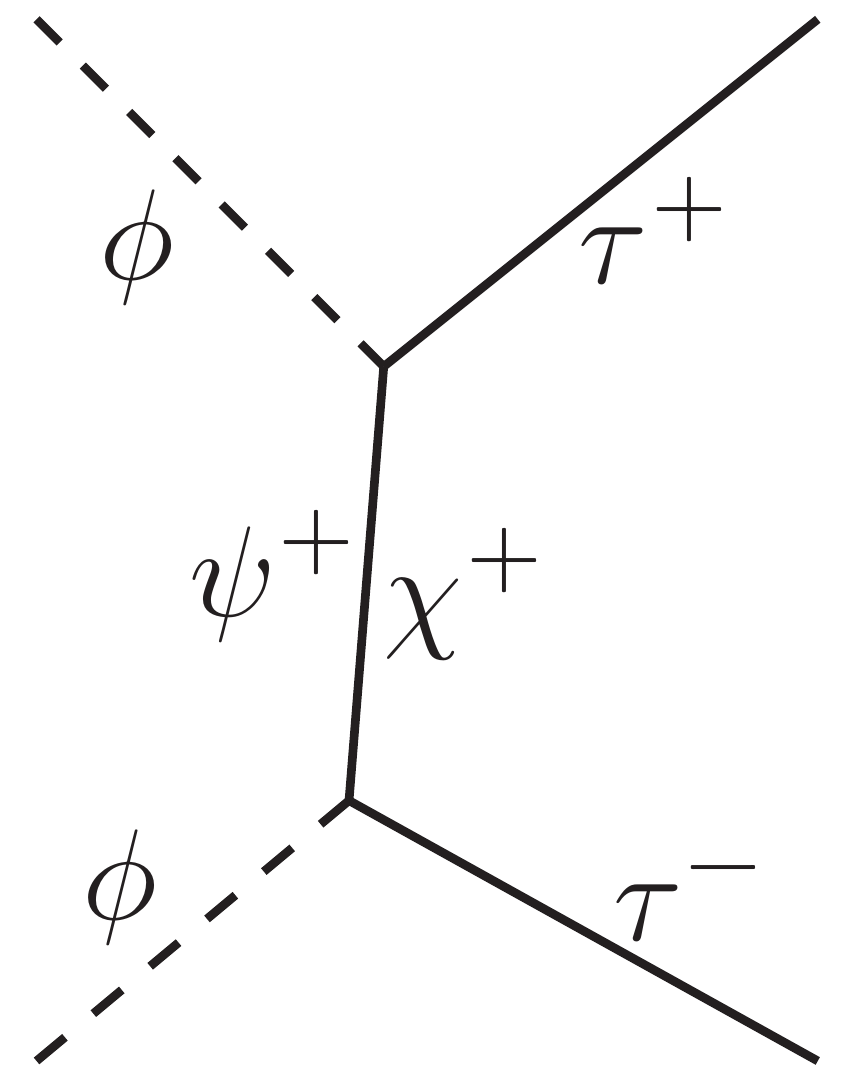}
  \caption{DM indirect search channel. Dark fermions appear in the propagator, facilitating the mixing dependence.}
 \label{diag:ann1}
\end{figure}

Apart from the discovery potential at the collider discussed above, the dark sector mixing can affect other DM detection possibilities as well. For the leptophilic scalar DM discussed here, the dominant indirect detection channel is $\phi \phi\to~\tau^+ \tau^-$ (\cref{diag:ann1}). We discussed the Fermi-LAT constraints on parameter space for the scalar DM and lepton doublet interaction in  Ref. \parencite[p.7]{Chakraborti:2019fnz}. There, we showed that most of the parameter region is allowed by experimental bound, except for a small region at low DM mass. In $m_{\phi} \lesssim$ 200~GeV, the region with large Yukawa coupling ($y_\tau \gtrsim 2.0$) is excluded by Fermi-LAT limits and this bound becomes more stringent if one considers small $\Delta m$.

In this work, the velocity-averaged annihilation cross section, i.e., $\ev{\sigma v}_{\tau\tau}$ will show dependence on the mixing parameter as the charged dark leptons appear in the $t$-channel propagator. In \cref{ind_det}, we show that for fixed $\delta m$'s and Yukawa couplings, a finite mixing can bring $\ev{\sigma v}_{\tau\tau}$ below the Fermi-LAT bounds which would otherwise be above the limits in \cite{Chakraborti:2019fnz}.

\begin{table}[!ht]
 \centering
 {\setlength{\tabcolsep}{1em}
 \begin{tabular}
  { c | c | c | c | l | l | l | c }
  \hline
      &\multicolumn{1}{ c }{$m_\phi$}
               & \multicolumn{1}{ c }{$m_{\psi^\pm}$}
                       & $m_\chi$
                              & \multicolumn{1}{ c }{\multirow{2}*{$y_\tau^D$}}
                                     & \multicolumn{1}{ c }{\multirow{2}*{$y_\tau^S$}}
                                            & \multicolumn{1}{ c |}{\multirow{2}*{$s_\alpha$}}
                                                   & $\ev{\sigma v}_{\tau\tau}$ \\
      & \multicolumn{3}{ c|}{[ GeV ]}
                              & \multicolumn{1}{ c }{}
                                     & \multicolumn{1}{ c }{}
                                            & \multicolumn{1}{ c |}{}
                                                   & [ cm$^3$/s ]          \\
  \hline
  \hline
  BP1 &  $100$ & $600$ &$110$ &$2.5$ &$0.0$ &$0.03$& $8.64\times 10^{-28}$ \\
  \hline
  BP2 &  $105$ & $600$ &$130$ &$2.0$ &$0.0$ &$0.45$& $6.23\times 10^{-27}$ \\
  \hline
  BP3 &  $125$ & $300$ &$140$ &$1.75$&$0.03$&$0.25$& $1.39\times 10^{-27}$ \\
  \hline
  BP4 &  $150$ & $400$ &$175$ &$2.5$ &$0.0$ &$0.27$& $5.98\times 10^{-27}$ \\
  \hline
 \end{tabular}
 }
 \caption{Relic density allowed benchmark points which are below the Fermi-LAT limit. Around $m_{\phi} \sim$ 100 GeV, the Fermi-LAT bound is $\sim 10^{-26}$ cm$^3$/s.}
 \label{ind_det}
\end{table}

The bound on the upper limit of Yukawa coupling can be relaxed for a finite dark sector mixing. It is obvious that due to having two charged dark fermions in this work, there are two $\Delta m (=\delta m/m_{\phi}$) parameters. On keeping one $\Delta m$ small and setting the other at a high value, it is possible to address both relic density and indirect search constraints. It becomes very interesting to apply this in low $m_{\phi}$ and small $\Delta m$ scenario, because as said previously, this region is typically above the limits in similar models.

The benchmark points in \cref{ind_det} are so chosen that for each point, DM-SM interaction takes place predominantly through the mixing. The relic density, on the other hand, is satisfied through various coannihilation and mediator annihilation channels which also have a strong mixing dependence. We observe here that for low DM mass of around 100 GeV, $\ev{\sigma v}_{\tau\tau}$ remains below the Fermi-LAT limit even for coupling as high as $y_\tau^D \sim 2.5$ and one of the $\Delta m$'s sufficiently small.

\subsection{Combined parameter region}
From the discussions above, the effect of dark matter mixing is evident in all the signatures of this model. In \cref{sec:lhc} we discussed the collider signatures for the \(3\tau + E^{miss}_T\) and \(\ell\,\tau + E^{miss}_T\) channels through different kinematic distributions, whereas \cref{sec:det} gives us the effect of mixing in the indirect search experiments. Here, in this section, we present a combined analysis where we take into account all the constraints together and see the prospects of detection of such signatures in the current and future experimental facilities.

Following the definition of the {\em fiducial phase space} in Ref. \cite{Aaboud:2017qkn}, the criteria for the same are the selection of {\em three $\tau$-jets (one light charged lepton and one $\tau$-jet) plus non-zero $E^{miss}_T$} for \(3\tau + E^{miss}_T\) (\(\ell\,\tau + E^{miss}_T\)) channel along with the cuts mentioned in \cref{tab:fid}. The {\em fiducial cross section} is defined as
\begin{align}
  \sigma_{\rm fid}
  =
  \frac{N - B}{\epsilon \cdot {\mathscr L}}
  \label{eq:fid}
\end{align}
where ${\mathscr L}$ is the integrated luminosity, $N$ is the expected number of observed events, $B$ is the estimated number of background events and $\epsilon$ is the efficiency of the detector to separate the particular signal events from the backgrounds. $B$ includes the background events from the final states given in \cref{tab:bkgs}. Since the SM sources of \(3\tau + E^{miss}_T\) (\(\ell\,\tau + E^{miss}_T\)) channel are also taken among the background processes, the numerator of \cref{eq:fid} gives us the excess number of events for the new physics signal. Thus, we can write, $N - B = \epsilon \cdot \sigma_{\rm fid} \cdot {\mathscr L} = S$. We estimated $\epsilon$ in our simulation as the ratio of the number of signal events passing the selection requirements at detector level to those passing the fiducial selection criteria listed in \cref{tab:fid} at the parton level. Hence, using the $\tau$-tagging efficiency and faking rate mentioned earlier, we calculate $N$, using \cref{eq:fid}, which gives us the estimated number events necessary to see the difference between the distributions of different benchmark points at a particular integrated luminosity.

\begin{table}[!ht]
 \centering
 {\setlength{\tabcolsep}{1em}
 \begin{tabular}
 { c | c@{} @{}l | r@{} @{}l | r | c }
 \hline
 & \multicolumn{2}{ c | }{backgrounds}            & \multicolumn{2}{ c | }{\makecell[c]{cross section \\ [1pt] [ pb ]}}
                                                                   &  events & total events   \\
 \hline
 \hline
 \multirow{4}*{\makecell[c]{\(3\tau +\) \\ [1em] \(E^{miss}_T\)}}
 &
 $t \widebar t$&$\to \tau\,2j\,\nu\,b \widebar b$ &    $66.$&$275$ &   $397$ & \multirow{4}*{$33069$} \\
 \cline{2-6}
 &
 $WW$&$\to \tau\,2j\,\nu$                         &    $10.$&$073$ &    $60$ &  \\
 \cline{2-6}
 &
 $WZ (\gamma^*)$&$\to \tau\,2j\,\nu$              &     $0.$&$125$ & $29908$ &  \\
 \cline{2-6}
 &
 $WZ (\gamma^*)$&$\to 3\tau\,\nu$                 &  $4984.$&$69$  &  $2704$ &  \\
 \hline
 \multirow{7}*{\makecell[c]{\(\ell\,\tau +\) \\ [1em] \(E^{miss}_T\)}}
 &
 $t \widebar t$&$\to \ell\,\tau\,2\nu\,b \widebar b$ &    $33.$&$14$  &  $19884$ & \multirow{7}*{$327626$} \\
 \cline{2-6}
 &
 $t \widebar t$&$\to \ell\,2j\,\nu\,b \widebar b$    &   $199.$&$194$ &   $3983$ &  \\
 \cline{2-6}
 &
 $WW$&$\to \ell\,\tau\,2\nu$                         &     $5.$&$026$ &   $3015$ &  \\
 \cline{2-6}
 &
 $WW$&$\to \ell\,2j\,\nu$                            &    $30.$&$239$ &    $604$ &  \\
 \cline{2-6}
 &
 $WZ (\gamma^*)$&$\to \ell\,2\tau\,\nu$              &     $0.$&$376$ &    $451$ &  \\
 \cline{2-6}
 &
 $WZ (\gamma^*)$&$\to \ell\,2j\,\nu$                 & $14972.$&$851$ & $299457$ &  \\
 \cline{2-6}
 &
 $WZ (\gamma^*)$&$\to 2\ell\,\tau\,\nu$              &     $0.$&$388$ &    $232$ &  \\
 \hline
 \end{tabular}
 }
 \caption{Estimate of the background events for the integrated luminosity, ${\mathscr L} = 100~\text{fb}^{-1}$.}
 \label{tab:bkgs}
\end{table}

\begin{figure}[!ht]
 \centering
 \includegraphics[width=0.48\textwidth]{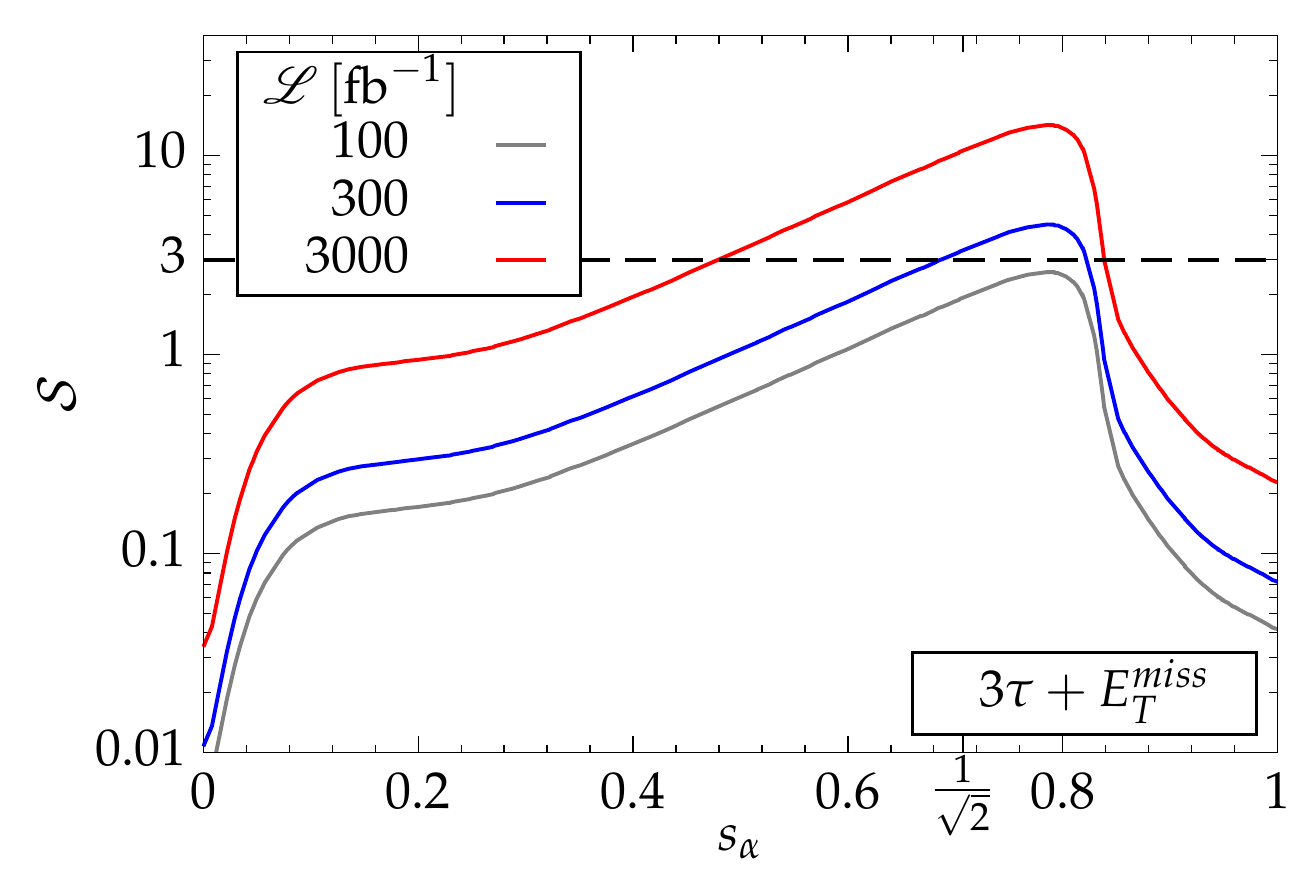}
 \includegraphics[width=0.48\textwidth]{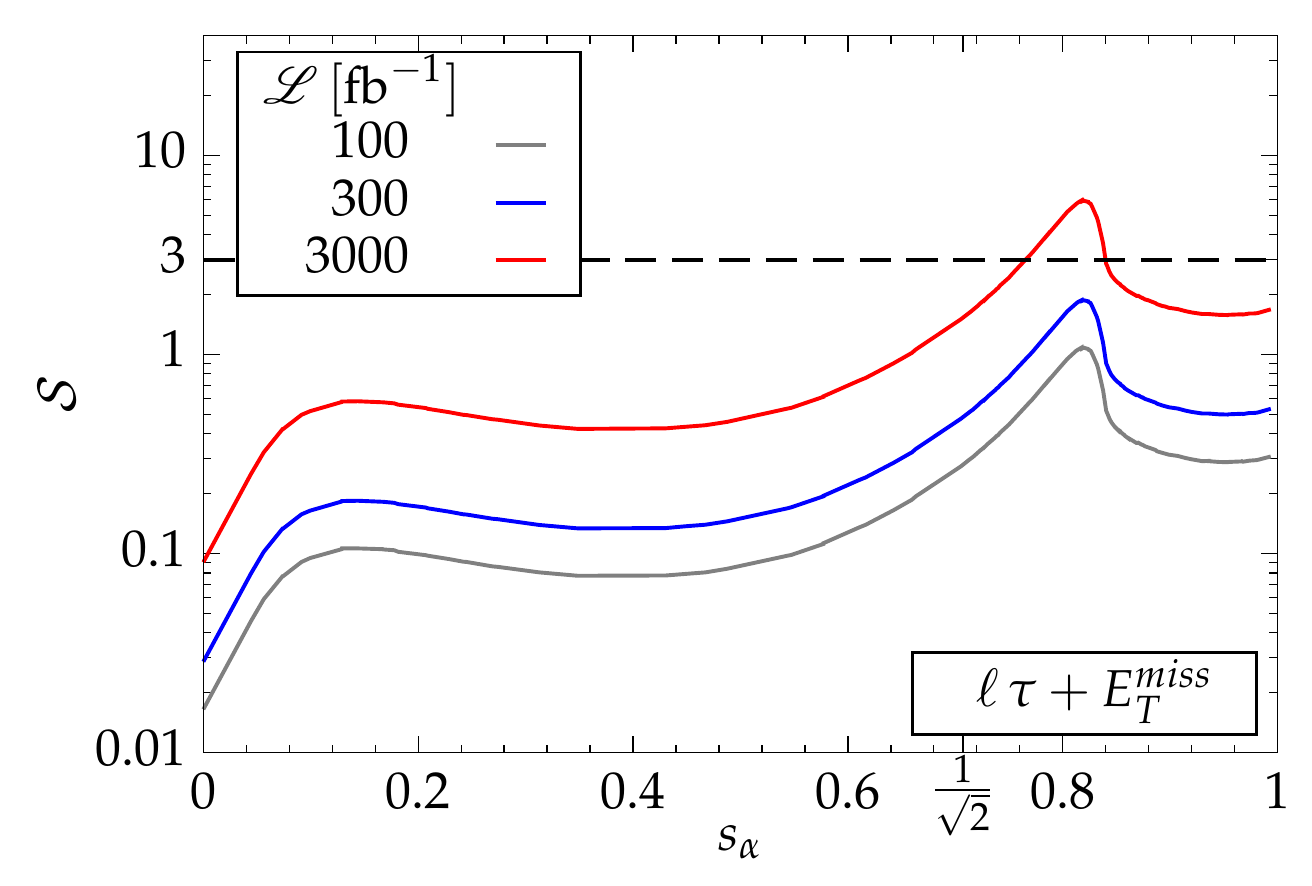}
 \caption{The variation of the statistical significance, ${\cal S}$ with the mixing parameter $s_\alpha$ for fixed values of other free parameters [~$(m_\phi, m_{\psi^\pm}, m_\chi) = (120, 420, 150)$~GeV and $(y_\tau^D, y_\tau^S) = (0.4, 0.21)$~]. ${\cal S} = \sqrt{2\left[(S+B)\ln(1+S/B) - S\right]}$, where $S$ and $B$ respectively are the number of expected signal and background events at a particular integrated luminosity ${\mathscr L}$.}
 \label{fig:signif}
\end{figure}

Taking all the considerations mentioned above, in \cref{fig:signif}, we show the variation of the statistical significance ${\cal S}$ as function of the mixing angle $\alpha$ for fixed values of other parameters, namely, $(m_\phi, m_{\psi^\pm}, m_\chi) = (120, 420, 150)$~GeV and $(y_\tau^D, y_\tau^S) = (0.4, 0.21)$. We have chosen three different values of ${\mathscr L} = 100, 300~\text{fb}^{-1}$ and $3~\text{ab}^{-1}$ for our study. We observe that ${\cal S}$ increases as $s_\alpha$ approaches $1/\sqrt{2}$. From \cref{fig:signif}, it is also evident that the \(3\tau + E^{miss}_T\) channel gives us better chance of detecting the mixing signatures for a larger region of parameter space than the \(\ell\,\tau + E^{miss}_T\) channel.

\begin{figure}[!ht]
 \centering
 \includegraphics[width=0.48\textwidth]{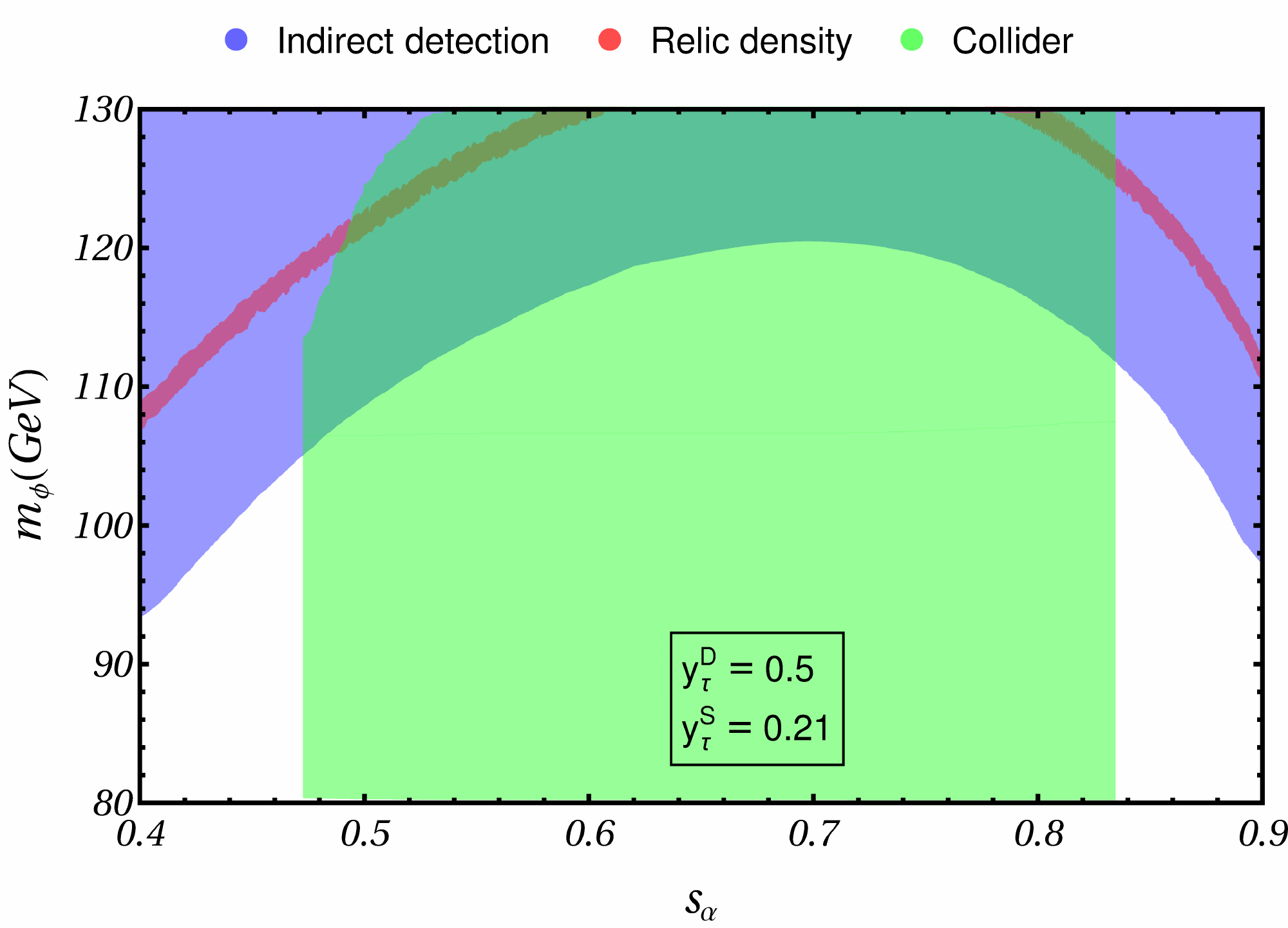}
 \includegraphics[width=0.48\textwidth]{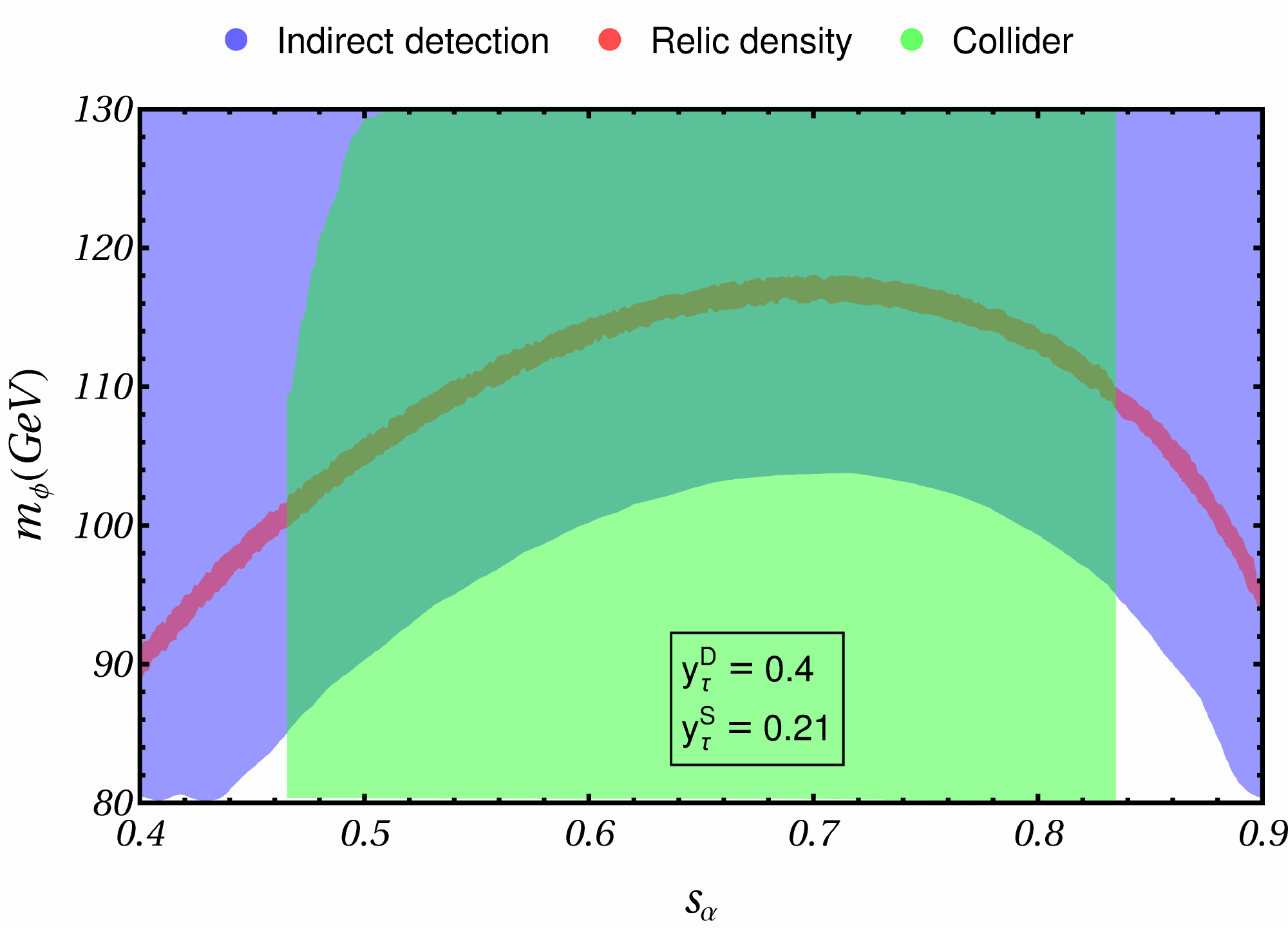}
 \caption{Parameter region combining all the experimental constraints. The blue shaded region denotes the region allowed by Fermi-LAT bounds, the red shaded area represents the relic density and the green shaded region gives the LHC limits for \(3\tau + E^{miss}_T\) channel at ${\mathscr L} = 3$~ab$^{-1}$, corresponding to ${\cal S} \ge 3$. Moreover, the entire parameter space is allowed by the direct detection constraints. The charged dark lepton masses are fixed at $(m_{\psi^\pm}, m_\chi) = (420, 150)$~GeV, whereas $m_{\psi^0}$ varies with $s_\alpha$ and can be calculated from \cref{eq:unphys_mass}.}
 \label{fig:param}
\end{figure}

Therefore, we perform a combined analysis and show the $s_\alpha$ vs $m_\phi$ correlation in \cref{fig:param}, taking into account all the experimental constraints. The shaded regions represent the allowed parameter space, with the blue, red and green shade corresponding to indirect detection, relic density and collider limits respectively. Here, we have shown the collider limits for \(3\tau + E^{miss}_T\) channel at ${\mathscr L} = 3~\text{ab}^{-1}$. This corresponds to ${\cal S} \ge 3$ for $0.46 \le s_\alpha \le 0.83$, as seen for the green shaded area. However, for smaller $s_\alpha$ in the above window, one can see that very large $m_\phi$ is not allowed. The low value of ${\cal S}$ in this region is consistent with that of \cref{fig:signif}.

One can set limits on the DM mass and the mixing angle by observing the region of overlap between all the shaded areas. However, with proper choice of the couplings, the overlap can be tuned. We show this in the two plots in \cref{fig:param}. Therefore, it can be concluded from these figures that the constraints on the mixing angle from the experimental limits can be tuned primarily with the proper choice of other parameters.

%
\section{Conclusion}
\label{sec:conc}
We have proposed a viable model with a leptophilic singlet scalar WIMP DM. A $\mathbb{Z}_2$ symmetry renders the stability to the dark sector. Apart from the DM candidate, the dark sector consists of a $SU(2)_L$ doublet and a singlet fermion. The presence of only a $SU(2)_L$ doublet in the dark sector which interacts through gauge as well as Yukawa couplings to the SM, adds new annihilation channels in the relic density calculation. However, in that case, one cannot distinguish one component of the doublet from the other by any means because of the degeneracy in their masses. Even if the degeneracy is lifted by the introduction of a new scale of EWSB through the extension of the scalar sector, both the doublet components interact through the same couplings and channels. This poses a serious problem in their distinction from one another. Introduction of a dark singlet in this scenario yields interesting features in the phenomenology. This is due to the fact that now, based on the electric charge of the new singlet, one component of the doublet will mix with it while the other component remains independent of mixing. Not only this additional degree of freedom automatically lifts the mass degeneracy of the dark fermions without making it an ad hoc proposition, but through the dark sector mixing, we can distinguish their effects in an experimental setup. Hence, on one hand, our model revives the simplest model of scalar singlet DM from the clutches of the stringent bounds of DM search experiments. On the other hand, it opens new search possibilities in the controlled environment of non-collider as well as collider based experiments. We conclude our observation and inferences in the following:

(1) For a better understanding of the mixing effects, in the relic density calculation, we choose the parameter region where the coannihilation channels are dominant. It is well known in the literature that the presence of coannihilation channels can boost up the relic density without adding to DM direct searches. For a similar model in Ref. \cite{Chakraborti:2019fnz} with leptophilic scalar DM and a fermionic doublet partner, we explored the viable parameter region thoroughly. In the present work, we show that in comparison with the previous study, here it is possible to relax the parameter region by a few orders of magnitude for appropriate tuning of the mixing parameter. This is due to the fact that as the dark fermions can non-trivially coannihilate now, effectively more channels are added to the total annihilation cross section. To compensate for this increase, we showed in the analysis that for larger mixing in the coannihilation regime, one needs to have a smaller coupling in order to be relic density allowed. This makes a larger parameter space viable over the full range of the mixing compared to the previous work.

(2) We show in the analysis that mixing can be a very useful tool in discriminating between the dark sector particles of different isospins. This is because of the fact that for the two extrema of the mixing angle, one of the dark partners is purely a singlet while the other one remains a pure $SU(2)_L$ doublet. For the intermediate values of the mixings however, it is obvious that the charged dark partners are mixed states. For very low or high mixing, coannihilation of these two dark fermions with DM will be substantially different from each other. This primarily because of the channels involving $W$ boson, which are available for the $SU(2)_L$ doublet dark partner for very low mixing and the singlet dark fermion for very high mixing. In our analysis, we discuss with correlation plots how these Gauge couplings help to clearly demarcate the parameter region w.r.t the contribution of the singlet and doublet dark partners towards total DM annihilation. However, as the mixing increases, the relative contribution of these dark sector particles accordingly vary. In the mixed scenario, it is interesting to observe how the viable parameter region evolves when the other parameters are fixed and it is only up to the mixing parameter to dictate the contribution of various dark sector coannihilation channels. 

(3) The mixing can directly affect various DM search prospects, e.g., indirect detection and collider searches. In indirect detection, the velocity-averaged annihilation cross section has a dependence on mixing due to having mixed states in the propagator. We show in our analysis that for low DM mass, it is possible to relax the existing bound on the upper limit of Yukawa coupling in the presence of mixing.

(4) From the observations of the kinematic distribution of various observables for the \(3\tau + E^{miss}_T\) and \(\ell\,\tau + E^{miss}_T\) channels, we conclude that one can clearly distinguish the effects of the mixing parameter that remains unaffected by the change in other free parameters of the model. That this gradual change in the peak and tail positions of the kinematic distributions with the variations of mixing is independent of other free parameters, was further established by changing the mixing parameter and keeping the rest of the independent parameters fixed. We can ensure the presence of the mixing parameter between the dark sector particles of the theory by looking at the peak and tail positions these distributions. That this feature of mixing is not limited to the Dirac fermions only can be concluded from the other studies in the literature \cite{Cirelli:2005uq,Cohen:2011ec}.

(5) We have studied the statistical significance for the above two final states at 13 TeV LHC. Between the two, the \(3\tau + E^{miss}_T\) channel can probe the signal for a larger range of mixing parameter. While it is not possible to observe the signal at the present LHC luminosity, we have shown with the numerical estimates that the signal can be observed at higher luminosities of 300 $\rm fb^{-1}$ and 3 $\rm ab^{-1}$, which are projected to be achievable at LHC within a few years.

(6) Lastly, we have done a scan of the parameter space in DM mass vs mixing angle plane, combining all the experimental constraints coming from DM relic density, direct detection and indirect detection as well as the collider constraints. For fixed values of the dark sector masses, it is indeed possible to exclude a portion of parameter space, but this can be tuned with the proper choice of DM-SM couplings. This is due to the fact that the effects of the couplings and dark sector masses are non-trivial on the various phenomenological aspects of this feature-rich model. Therefore, it is not straightforward to obtain such exclusion limits on mixing angle without taking into account the effect of the other parameters.

We conclude this article with the assertion that the mixing between a singlet and a doublet dark sector fields can turn the table in favour of a quintessential scalar singlet DM model. It evades the stringent experimental bounds from the DM detection experiments as well as presents new opportunities for its detection in the lab.

\acknowledgments

The authors thank Sabine Kraml for useful suggestion. RI thanks the SERB-DST, India for the research grant EMR/2015/000333. SC acknowledges MHRD, Government of India for research fellowship.

\printbibliography
\end{document}